\documentclass[11pt, a4paper, copyright]{perplexity}

\usepackage[authoryear, sort&compress, round]{natbib}
\usepackage{xspace}
\usepackage{graphicx}  
\usepackage{float}                                               
\usepackage{adjustbox}
\usepackage{tablefootnote}
\usepackage{svg}
\usepackage{array}

\usepackage{tcolorbox}

\newcolumntype{L}[1]{>{\raggedright\arraybackslash}p{#1}}
\newcommand{\ours}{Q2D-Web\xspace}
\AtBeginDocument{%
}

\begin{document}
\uselogo{}

\title{Q2D-Web: A Large-Scale Benchmark for Retrieval in Agentic RAG Systems}

\author[1]{Maximilian Schall}
\author[1]{Sedigheh Eslami}
\author[1]{Markus Krimmel}
\author[1]{Antoine Chaffin}
\author[1]{Louis Milliken}
\author[1]{Bo Wang}
\author[1]{Denis Bykov}

\affil{Perplexity AI}

\begin{abstract}
Evaluating first-stage retrievers in large-scale production RAG requires a benchmark that pairs a large-scale corpus with a large set of agent-reformulated search queries based on real user queries and their conversation threads, and that labels many relevant documents per query. No existing public benchmark evaluates this setting: large-scale collections typically provide only a small number of evaluation queries, whereas benchmarks with many queries generally contain only millions of documents. Moreover, most benchmarks assess human-written queries, while the first-stage retrievers in agentic RAG pipelines serve machine-written reformulations whose distribution differs from human search behavior. To overcome these evaluation gaps, we introduce \ours{} (Query2Doc-Web), a large-scale agentic retrieval benchmark consisting of a $\sim$190M-document web corpus and $\sim$70k agentic search queries in ten languages, reformulated from real-world user queries in production systems. \ours{} provides three sets of fixed relevance judgments: agent citations, production rankings, and a combined set that unions both signals and adds LLM-based judgments of unlabeled pooled documents to reduce false negatives. We benchmark 13 retrievers including lexical, dense, and late-interaction models and find that their relative ordering is largely insensitive to the choice of judgment set, while diverging substantially across topical domains, query languages, and query types. To enable fast evaluation, we also study subcorpus sampling as an approximation to full-corpus evaluations. Retaining a third of the corpus, selected by reciprocal rank fusion over pooled retriever runs, preserves the full-corpus model ranking under the combined judgments while raising absolute Recall@1000 only by 4--7 points. The public leaderboard is accessible under: \url{https://huggingface.co/spaces/perplexity-ai/q2d-web-leaderboard}.
\end{abstract}

\maketitle

\section{Introduction}
\label{sec:intro}
Agentic retrieval-augmented generation (RAG) relies on retrieval to identify evidence for accurate, grounded responses~\citep{Lewis2020RAG, OpenAI2025DeepResearch, Xi2025DeepSearchSurvey}.
In a production web search scenario, retrieval is performed through a multi-stage pipeline over large and heterogeneous document collections: a first-stage retriever scans the full index and returns a candidate set that later stages rerank~\citep{Wang2011Cascade, Nogueira2019MultiStage}.
The first-stage retriever consequently bounds what the agent can read and ultimately cite.
As such, an evaluation of this component should test retrieval under realistic production-scale conditions, in which systems must rank relevant documents ahead of semantically similar but non-relevant alternatives~\citep{Reimers2021Curse}. Combining a large search space, many production queries, and deep relevance judgments lets us measure how much relevant evidence a retriever recovers among plausible alternatives.

Existing benchmarks do not support evaluation at such a large scale.
Commonly used collections, such as MS~MARCO~v1~\citep{Bajaj2016MSMARCO} and v2~\citep{craswell2021trec}, contain at most 12M documents.
MS~MARCO Web Search~\citep{Chen2024MSWS} increases the collection size to 100M documents, but includes only 9{,}374 test queries. Previous work~\citep{Voorhees2002Reliability} motivates the inclusion of larger query sets to reduce the retrieval-experiment error and make system comparisons less sensitive to the particular queries sampled. Evaluation should therefore use a large, topically diverse collection of queries.
Furthermore, agentic benchmarks such as BrowseComp-Plus~\citep{Chen2025BrowseCompPlus}, although providing a fixed corpus with human-verified evidence documents, contain only 830 research queries.

Production queries issued to the retriever are also reformulated versions of the user's natural queries. More specifically, the agent reformulates, decomposes, and iteratively revises the original user query using conversational context before each retrieval step~\citep{Xi2025DeepSearchSurvey, Jin2025SearchR1}. Retrieval effectiveness is sensitive to how a query is formulated. Even reformulations that preserve the intent reduce the nDCG@10 of retrieval pipelines by about 20\% on average~\citep{Penha2022QueryVariations}.

\begin{figure}
    \centering
    \includegraphics[width=\linewidth]{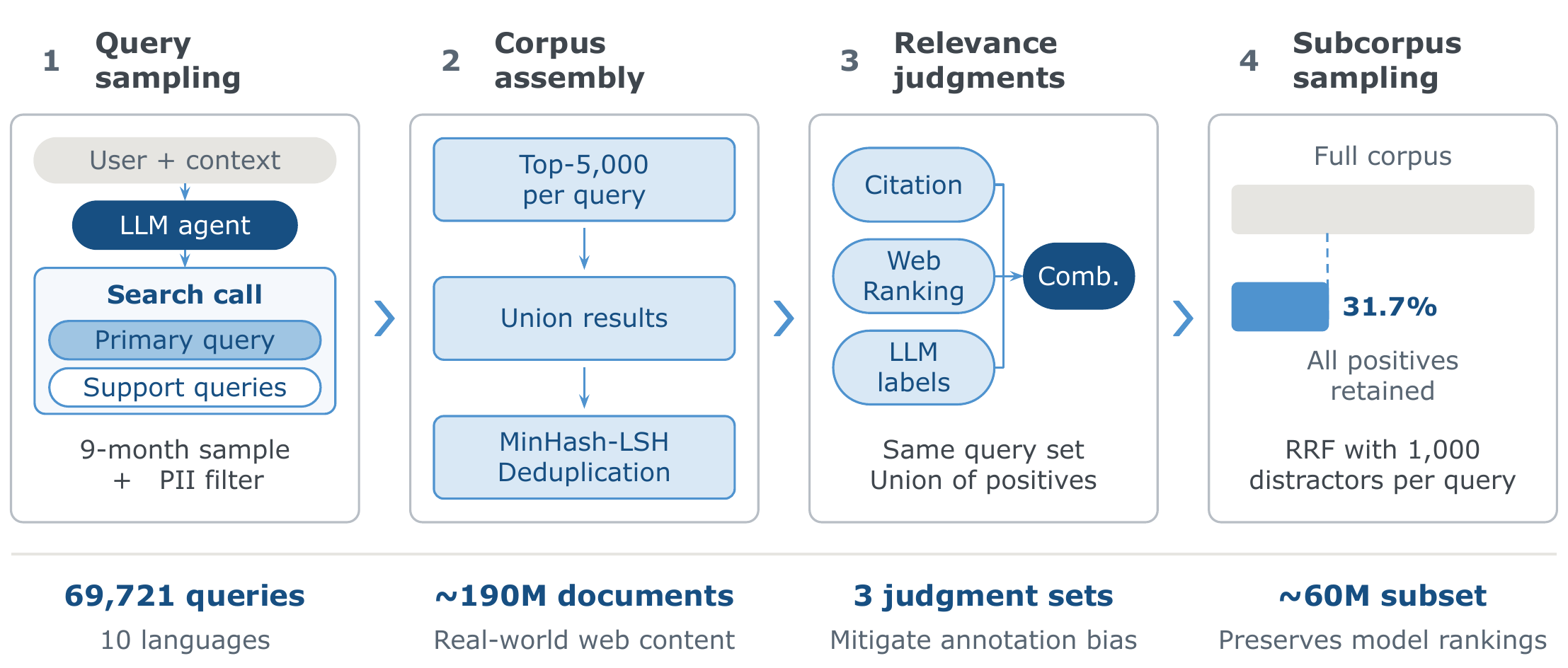}
    \caption{Q2D-Web construction pipeline. We sample queries from agent search calls, assemble and deduplicate the corpus, construct three relevance-judgment sets, and sample an evaluation subcorpus while retaining all labeled positives.}
    \label{fig:pipeline}
\end{figure}

Beyond these limitations in scale and query type, existing collections also provide shallow relevance supervision.
MS~MARCO Web Search, for instance, assigns a relevance label to only one clicked document per query, leaving potentially relevant but unclicked documents unlabeled.
Because unjudged documents are conventionally treated as non-relevant, a system that retrieves relevant but unlabeled documents receives a lower score~\citep{Buckley2004Incomplete}.
Such false negatives are common under sparse labels: on MS~MARCO, 70\% of manually inspected top-retrieved but unlabeled passages were in fact relevant~\citep{Qu2021RocketQA}, and preference judgments rated the top results of a modern ranker above the officially judged relevant items~\citep{Arabzadeh2022ShallowPooling}.
These findings motivate deeper relevance assessment, which judges more retrieved documents per query and reduces the risk that relevant results are unlabeled, mitigating the possibility of false negatives in the corpus.

To enable a faithful evaluation of such first-stage retrievers in large-scale production RAG systems, we introduce \ours{} (Query2Doc-Web), a benchmark that combines a $\sim$190M-document web corpus with $\sim$70k agent-reformulated queries in ten languages, sampled from nine months of PII-free production traffic.
To limit false negatives without inheriting the bias of any single label source, we construct three judgment sets, derived from agent citations, production rankings, and LLM judgments of previously unjudged candidates.
\autoref{fig:pipeline} summarizes the construction pipeline, from query sampling and corpus assembly to relevance assessment and subcorpus selection.
To the best of our knowledge, \ours{} is the first benchmark to jointly provide a large-scale corpus, a large set of production-sampled agent-reformulated queries, and deep reusable relevance judgments. \autoref{tab:benchmarks} shows how \ours{} differs from currently available benchmarks in scale, query type, and label coverage.

Since evaluations on the full \ours{} corpus are costly, we also systematically study subcorpus sampling as an approximation to full-corpus evaluation. 
We find that retaining a third of the corpus, selected by reciprocal rank fusion (RRF) score over pooled retriever runs, preserves the full-corpus ranking of systems and approximates the absolute scores. 
This enables a two-stage protocol for evaluation: a new retriever can first be evaluated on the sampled subcorpus, and promising retrievers can then be assessed against the full corpus.

We evaluate lexical, dense, and late-interaction retrievers on the full corpus as well as the subsampled version. To assess performance in a production environment, where the retriever selects a candidate pool for downstream re-ranking, we use a deep-recall protocol. We focus on Recall@1000 as our primary metric and report Recall@100 and nDCG@10 to assess retrieval performance at shallower cutoffs.
Beyond aggregate effectiveness, we find that the number of relevant documents uniquely recovered by a retriever is not ordered by its aggregate recall: the weakest retriever by Recall@1000 contributes the largest set of relevant documents that no other retriever recovers, so retriever family matters independently of scale. We further characterize unique contributions, hard positives, and false negatives.

We argue that publicly releasing \ours{} would erode its validity over time.
Public benchmarks often end up in training data, intentionally or unintentionally through derivative datasets, and models trained on them obtain inflated scores~\citep{Sainz2023Contamination, Matton2024CodeLeakage}.
Because most training corpora are not public, such contamination is difficult to rule out after release~\citep{Oren2024Proving}.
We therefore keep the corpus, queries, and judgments private and operate a public leaderboard that evaluates open-weight retrievers: \url{https://huggingface.co/spaces/perplexity-ai/q2d-web-leaderboard}

\vspace{2mm}
In summary, our contributions are as follows:
\begin{itemize}
  \item \textbf{A large-scale agentic retrieval benchmark dataset}
 containing a $\sim$190M-document web corpus with $\sim$70k agent-reformulated queries in ten languages, sampled from nine months of production traffic. Its queries are the machine-written reformulations that first-stage retrievers serve in deployed production environments.
  \item \textbf{Multi-source relevance judgments}: three reusable judgment sets over the same queries and corpus, which reduce dependence on any single source of relevance labels.
  \item \textbf{Large-scale subcorpus sampling} demonstrating, for the first time, evaluation-preserving sampling on a large-scale corpus with about 190M documents. The resulting subcorpora support efficient benchmarking and rapid iteration while reproducing the full-corpus system ordering under the Combined label set.
  \item \textbf{Benchmarking 13 modern retrievers} spanning lexical, dense, and late-interaction architectures under a deep-recall protocol. We find that different models achieve the highest Recall@1000 and nDCG@10 (\autoref{sec:main_results}). All tested neural retrievers have lower recall on supporting queries than on primary queries, while BM25 has slightly higher recall on supporting queries (\autoref{fig:primary-vs-supporting}). BM25 also has the lowest Recall@1000 overall, yet recovers the largest set of relevant documents that no other tested retriever recovers (\autoref{sec:unique-positives}).
\end{itemize}

\begin{table*}[t]
\caption{Public IR collections compared to \ours{}. \#Indexable units denotes the
number of items retrieved and ranked by a system. Test q.\ is the number of test
queries. Judg./q.\ is the mean number of positive relevance judgments per test
query; n/a indicates that we could not determine a comparable value. References
for all collections are given below the table.}
\label{tab:benchmarks}
\centering
\resizebox{\textwidth}{!}{%
\footnotesize
\setlength{\tabcolsep}{3.5pt}
\begin{tabular}{@{}l r r r r l l l@{}}
\toprule
\textbf{Collection} & \textbf{\#Indexable units} & \textbf{\#Lang.} &
\textbf{Test q.} & \textbf{Judg./q.} &
\textbf{Query origin} & \textbf{Relevance labels} & \textbf{Domain} \\
\midrule
TREC Web 2009--12
  & 1.04B pages & 10 & 50\,/\,yr & 63--137
  & search log & human, pooled, graded & web \\
TREC DL 2019--23
  & 138M passages & 1 & 43--82\,/\,yr & 67--1{,}315
  & search log & human, pooled, graded & web \\
TREC RAG 2024
  & 113.5M segments & 1 & 301 & n/a
  & search log & human + LLM, pooled & web \\
MS~MARCO Web
  & 100.9M documents & 93 & 9{,}374 & 1
  & search log & clicks & web \\
BioASQ
  & 14.9M articles & 1 & 500 & 4.7
  & domain experts & human & biomedical \\
MS~MARCO Passage
  & 8.8M passages & 1 & 6{,}980 & $\sim$1
  & search log & human, sparse & web \\
FEVER
  & 5.4M articles & 1 & 19{,}998 & n/a
  & crowdworker claims & human & Wikipedia \\
HotpotQA
  & 5.2M paragraphs & 1 & 7{,}405 & $\sim$2
  & crowdworkers & human & Wikipedia \\
AIR-Bench 24.05
  & 1.0M passages ($\times$69) & 13 & 102K & n/a
  & LLM-generated & LLM judge & 9 domains \\
BrowseComp-Plus
  & 100{,}195 documents & 1 & 830 & 6.1
  & human trainers & human-verified; gold-answer subset
  & web (curated) \\
\midrule
\ours{}
  & 190M documents & 10 & 70K & 88.5
  & LLM agents
  & \textsc{Citation} + \textsc{Web Ranking} + LLM
  & web \\
\bottomrule
\end{tabular}%
}

\vspace{1mm}
\begin{minipage}{\textwidth}
\footnotesize\raggedright
Sources: TREC Web 2009--12~\citep{ClueWeb09};
TREC DL 2019--23~\citep{Craswell2019TRECDL19, Craswell2023TRECDL23};
TREC RAG 2024~\citep{Upadhyay2024LargeScaleLLMAssess, Pradeep2024Ragnarok};
MS~MARCO Web Search~\citep{Chen2024MSWS};
BioASQ~\citep{Tsatsaronis2015BioASQ, Thakur2021BEIR};
MS~MARCO Passage Ranking~\citep{Bajaj2016MSMARCO};
FEVER~\citep{Thorne2018FEVER};
HotpotQA~\citep{Yang2018HotpotQA};
AIR-Bench 24.05~\citep{Chen2025AIRBench};
BrowseComp-Plus~\citep{Chen2025BrowseCompPlus}.
\end{minipage}
\end{table*}

\section{Related Work}
\label{sec:related}
\paragraph{IR Benchmarks.}
Information retrieval (IR) evaluation has long relied on benchmarks of limited scale, in both corpus size and query set size.
MS~MARCO~\citep{Bajaj2016MSMARCO} and the TREC Deep Learning Track~\citep{Craswell2019TRECDL19, Craswell2023TRECDL23} provide the most widely used passage- and document-ranking benchmarks, with document corpora up to 11.9M documents, e.g., in MS~MARCO~v2. BEIR~\citep{Thakur2021BEIR} broadens coverage to 18 domain/task combinations, but individual tasks remain small in both dimensions.
Corpus sizes span 3.6K to 5.4M documents, apart from BioASQ at 14.9M, and test sets 49 to 7{,}405 queries,  apart from CQADupStack and Quora, which contain 13{,}145 and 10{,}000 test queries, respectively.
Large-scale benchmarks partially address corpus scale with collections that are orders of magnitude larger but still suffer from limited numbers of queries.
The ClueWeb family~\citep{ClueWeb09, ClueWeb12, Overwijk2022CW22} provides crawls of 50M to 10B documents, but the TREC Web Track provides only 50 pooled, human-judged queries per year (2009--2012), fewer than one judged query per 10M documents.
MS~MARCO Web Search~\citep{Chen2024MSWS} builds the largest public collection with web click-derived labels, sampling 100M documents from ClueWeb22 and providing 9{,}374 labeled test queries with only one clicked document each.
\ours{} provides $\sim$190M documents with $\sim$70k judged queries, more than seven times as many judged queries as MS~MARCO Web Search, and judges 88{.}5 positive documents per query rather than one.

\paragraph{Agentic Evaluation and Relevance Judgment.} 
A parallel line of work evaluates language models and agents on fact-seeking web tasks rather than corpus-level retrieval.
SimpleQA~\citep{Wei2024SimpleQA} contains 4{,}326 short-answer questions authored by human trainers, each verified by an independent second trainer and required to have a single, time-invariant answer.
BrowseComp~\citep{DBLP:journals/corr/abs-2504-12516} extends this design to 1{,}266 harder questions built by inverse authoring: annotators start from a known fact and add constraints until the answer cannot be found on the first page of five search-engine queries and cannot be solved by frontier models or by another annotator within ten minutes.
Neither defines a document corpus nor $(q, d)$ relevance judgments; correctness is graded by matching a single short answer against the live web.
BrowseComp-Plus~\citep{Chen2025BrowseCompPlus} closes this gap by re-annotating 830 verified BrowseComp queries against a fixed 100{,}195-document corpus, with an average of 6.1 human-verified evidence documents, 2.9 gold documents, and 76.3 mined hard negatives per query.
This supports standard IR metrics such as nDCG and Recall.
These benchmarks evaluate the browsing or research agent as a whole, scoring its final answer or its set of source-backed records rather than the first-stage retriever that supplies its evidence.

\ours{} isolates the retriever at web scale: it evaluates first-stage retrieval on the agent-reformulated queries that production pipelines issue, against a corpus roughly 1{,}900 times larger and a judged query set roughly 84 times larger. Where these benchmarks rely on a single label source, \ours{} provides three reusable judgment sets over the same queries and corpus, derived from agent citations, production rankings, and LLM judgments of previously unjudged candidates.

\paragraph{Efficient and Faithful Corpus Subsampling.}
Evaluating retrievers at scale is computationally demanding.
A full pass over hundreds of millions of documents is costly for any modern retriever: dense encoders now reach 4B to 8B parameters, late-interaction models emit one vector per token, and even sparse indexes require substantial infrastructure to build and serve.
To reduce this cost, \citet{Froebe2025CorpusSubsampling} propose to construct a reduced evaluation corpus by taking the top-100 pool of the runs that contributed to the original judgment pool, and show on nine TREC tracks that Pool$_{100}$ is statistically indistinguishable from full-corpus nDCG@10 in both system ranking and absolute score, at up to 1{,}000 times lower compute.
Their follow-up work applies the same strategy to build a learned-sparse retrieval benchmark on 11 TREC tasks~\citep{DBLP:conf/ecir/FrobeSRHMHLNVP26}.
CoRECT~\citep{DBLP:conf/ecir/CaspariDDFMG26} adopts the strategy to construct MS~MARCO~v2 subsets up to 100M passages, using reciprocal-rank fusion over the pooled TREC DL 2023 runs to mine hard distractors before backfilling with random documents.
\citet{Dinzinger2026BridgingSubsampled} identify a residual gap in this paradigm: because subsampled corpora omit the long tail of non-relevant background documents, subsampled evaluation systematically overestimates nDCG@k and Recall@k, and the gap grows with $k$.
They correct the metric analytically by fitting a Gaussian to query-to-non-relevant similarity scores.
Both operate at TREC scale with human judgments as the anchor; we subsample a $\sim$190M-document production index against tens of thousands of citation-labeled agentic queries, where qrels combine explicit LLM judgments with implicit relevance signals, yielding substantially more labeled documents per query.

\section{\ours{}}
\label{sec:dataset}
In this section, we detail the four-stage pipeline for creating \ours{}: query sampling, corpus assembly, relevance-judgment annotation, and subcorpus sampling for efficient evaluation.

\subsection{Queries}
\label{sec:queries}
\ours{} contains $\sim$70k LLM-reformulated queries, generated from $\sim$23k production searches collected over a nine-month window.
A \emph{search} is all of the web-search activity the agent performs in response to a single user message: one or more tool calls, each carrying one or more query strings. The agent generates every query from the user's message, the preceding conversation, and any results already retrieved within the same search; users do not enter search queries directly.
In production, each search contains one \emph{primary} query, which restates the user's intent, and zero or more \emph{support} queries, which the agent formulates more freely to produce alternative phrasings, retrieve background material, or reach adjacent entities. \autoref{fig:example-search} shows three such searches, in which the primary and support queries draw on conversational context that the user message alone does not contain.
A support query remains within the primary query's domain but differs in surface form and in the set of documents that satisfy it. We therefore treat each query as a separate retrieval problem with its own judgments.
 
\begin{figure}
    \centering
    \includegraphics[height=4.2cm]{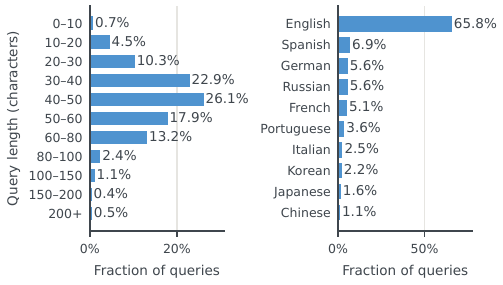}\hfill%
    \includegraphics[height=4.2cm]{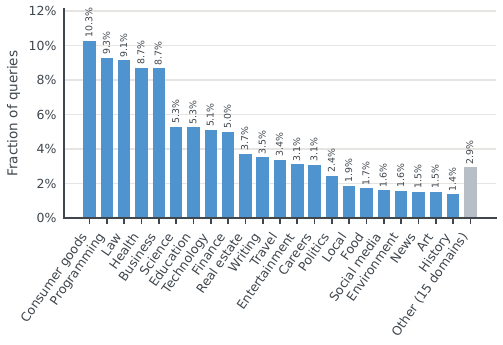}%
    \caption{Statistics of the \ours{} queries: character-length (left), language distributions (center), and domain distribution (right).}
    \label{fig:queries}
\end{figure}

Sampling is stratified by month across a nine-month deployment window to ensure a uniform distribution of search dates. Within each stratum, we match the production distribution over topical domain and language. We then discard exact duplicate queries and queries shorter than three characters. We also discard queries containing filtering operators, such as \texttt{site:}, and keep the ten most common languages in the dataset. Additionally, we apply personally identifiable information (PII) detection and exclude queries that contain PII. Filtering operates on individual queries rather than whole searches, so a search can lose its primary query while its support queries are retained. 
The resulting set contains $69{,}721$ queries: $12{,}365$ primary ($17.7\%$) and $57{,}356$ support ($82.3\%$) with approximately 4 support queries per primary query on average. \autoref{fig:queries} shows their character-length, language, and domain distributions. 

\begin{figure}[t]
\centering
\small
\begin{tcolorbox}[colback=gray!5, colframe=gray!40, boxrule=0.4pt,
                  left=4pt, right=4pt, top=3pt, bottom=3pt]
\textbf{User.} Our professor told us to read an article on privacy of the same journal, which one could it be?

\vspace{6pt}\hrule\vspace{6pt}

\textbf{Primary query.} \emph{seminal Harvard Law Review right to privacy article}

\vspace{6pt}\hrule\vspace{6pt}

\textbf{Support queries.}
\begin{itemize}[nosep, leftmargin=1.2em]
  \item \emph{Harvard Law Review privacy article professor reading assignment}
  \item \emph{most cited Harvard Law Review privacy articles}
\end{itemize}
\end{tcolorbox}

\vspace{2pt}

\begin{tcolorbox}[colback=gray!5, colframe=gray!40, boxrule=0.4pt,
                  left=4pt, right=4pt, top=3pt, bottom=3pt]
\textbf{User.} libri che capire meglio programmazione distribuita e cloud computing

\vspace{6pt}\hrule\vspace{6pt}

\textbf{Primary query.} \emph{libri che capire meglio programmazione distribuita e cloud computing}

\vspace{6pt}\hrule\vspace{6pt}

\textbf{Support queries.}
\begin{itemize}[nosep, leftmargin=1.2em]
  \item \emph{best cloud computing books}
  \item \emph{best books distributed systems}
  \item \emph{migliori libri cloud computing}
\end{itemize}
\end{tcolorbox}
\vspace{2pt}

\begin{tcolorbox}[colback=gray!5, colframe=gray!40, boxrule=0.4pt,
                  left=4pt, right=4pt, top=3pt, bottom=3pt]
\textbf{User.} list of samsung a series from highest to lowest in term of price and performance of the phone

\vspace{6pt}\hrule\vspace{6pt}

\textbf{Primary query.} \emph{Samsung Galaxy A series models 2025 ranked by price}

\vspace{6pt}\hrule\vspace{6pt}

\textbf{Support queries.}
\begin{itemize}[nosep, leftmargin=1.2em]
  \item \emph{current Samsung Galaxy A lineup specs comparison}
  \item \emph{Samsung A56 A36 A26 A16 A06 specs prices}
  \item \emph{current Samsung Galaxy A lineup highest to lowest}
\end{itemize}
\end{tcolorbox}
\vspace{-2mm}
\caption{Three example searches from \ours{}. Top: the agent resolves context from prior messages and reformulates around the right-to-privacy article. Middle: the user message is Italian for ``books to better understand distributed programming and cloud computing''. The agent formulates support queries in both languages for wider coverage. Bottom: support queries proceed in two hops, first establishing an overview, then resolving specific models.}
\label{fig:example-search}
\end{figure}

\subsection{Web Corpus}
\label{sec:corpus}
 
For each of the $\sim$70k queries, we retrieve the top 5{,}000 documents using a production retrieval system and take the union of these result sets. 
We then deduplicate this collection with MinHash--LSH~\citep{Broder1997SyntacticClustering}, clustering documents whose token $5$-gram sets have a Jaccard similarity of at least 0.975. This yields approximately 190M documents.
Within each cluster, we retain the longest document as the canonical representative and remap judgments associated with discarded documents to that representative. Each document consists of a title and a body. The character-length distribution of the corpus is reported in \autoref{fig:corpus_qrel}. On average, each document contains 13,169 characters.

The corpus is constructed from top-ranked retrieval results rather than a random sample of web documents. Consequently, it contains documents that a production retrieval system considered plausible matches for at least one dataset query. We argue that such documents are more likely to be challenging distractors for a retriever model than randomly sampled documents, which are often unrelated to the query.

This effect is strengthened by the query structure. A search contains at most one primary query and, in most cases, one or more support queries (\autoref{sec:queries}) that address the same user intent but retrieve different documents. Therefore, a document retrieved for one reformulated query can serve as a hard-negative candidate for a related query. The pooling of results across these related queries thus produces a corpus with a high concentration of plausible distractors without a separate stage of hard-negative mining. 

\subsection{Relevance-Judgment Passes}
\label{sec:qrels}
A judgment in \ours{} is a binary relevance label on a query--document pair. Labeling every pair is infeasible at this scale, but relevance in a pooled corpus overlaps across queries. A document that is a hard negative for one query could be a positive for a related one. We therefore derive relevance labels for the same queries and corpus from three different sources. Each source surfaces and judges relevant documents that the others miss,
which decreases false negatives. Using several sources reduces the risk of
source-specific bias, which favors retrievers similar to the labeling system.

\paragraph{Citation.}
A document is labeled relevant if an LLM agent cited it in its response. In this work, we assume that citation means the agent used the document to support a claim; therefore, we treat these cited documents as relevant to the query. However, the agent stops citing once a claim is supported, so a relevant document that is redundant with an already-cited source is left uncited and labeled non-relevant. \textsc{Citation} is therefore high-precision and low-recall by construction, and it is the only pass whose labels reflect downstream use of a document rather than an assessment of topical match.
\paragraph{Web Ranking.} This label set comprises up to the top 50 results per query ($43.1$ on average) from an internal retrieval system that employs BM25 and dense retrieval in the first stage, followed by cross-encoder reranking.
This pass labels relevant documents that citation behavior discards. Unlike \textsc{Citation}, whose labels are generated at the original search time, \textsc{Web Ranking} is computed against the corpus snapshot, so the two judgment sets might not overlap on documents that changed content over the nine-month window.
\paragraph{Combined+LLM-Judged.} This relevance set contains the union of the judgments from \textsc{Citation}, \textsc{Web Ranking}, and additional LLM-generated judgments. We use BM25~\citep{robertson2009bm25}, ColBERTv2~\citep{santhanam2022colbertv2}, the dense models arctic-l-v2~\citep{yu2024arctic}, bge-m3~\citep{chen2024bgem3}, e5-large-instruct~\citep{wang2024multilinguale5}, gte-multilingual-base~\citep{zhang2024mgte}, mxbai-large-v1~\citep{lee2024mxbai,li2023angle}, nomic-v1.5~\citep{nussbaum2024nomic}, and stella-400m-v5~\citep{zhang2025stella} to rank documents that have not been judged by either \textsc{Citation} or \textsc{Web Ranking}. \autoref{app:pooling-models} lists their
architectures, sizes, and input lengths. We select these retrievers by release date alone, keeping every model released before our cutoff of January 1, 2025, which keeps the pooling models separate from the evaluation models and avoids bias in favor of the evaluation models. We merge these rankings with reciprocal rank fusion, select the top 500 previously unjudged documents, and ask \textsc{deepseek-ai/DeepSeek-V4-Flash} to assign each query--document pair a binary relevance label (\autoref{fig:prompt}). 

\begin{figure}
    \centering
    \resizebox{\linewidth}{!}{%
        \includegraphics[height=4cm]{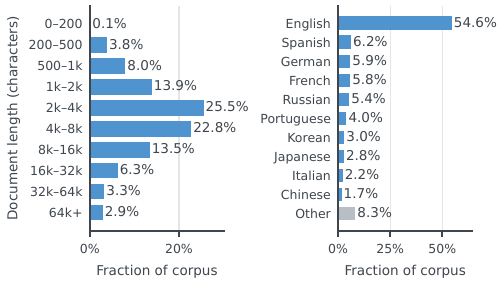}\hspace{2mm}%
        \includegraphics[height=4cm]{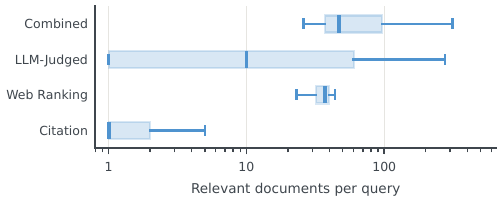}%
    }
    \caption{Corpus and judgment statistics. Left and center: character-length and language distributions of the corpus, where the Other category comprises 406 languages. Right: relevant documents per query for each judgment source and their union (Combined), showing the median, IQR, and 5th--95th percentiles on a logarithmic scale.}
    \label{fig:corpus_qrel}
\end{figure}

Each of these label sources carries its own distinct biases. \textsc{Citation} depends on the citing agent, its prompt, and the candidate set it retrieved. \textsc{Web Ranking} depends on the stack that produced it: the choice of first-stage retriever models and cross-encoder rerankers. An LLM-as-a-judge~\citep{gu2025surveyllmasajudge} has documented preferences of its own, for example, favoring fluent and long-form text. In order to reduce the risk of source-specific annotation bias, \ours{} includes all three relevance label sets. Each set comprises binary labels for query--document pairs and may mark multiple documents as relevant for a given query. \autoref{fig:corpus_qrel} gives an overview of the distribution. We evaluate every model separately using each label set and discuss the results in \autoref{sec:main_results}.

\subsection{Subcorpus Sampling}
\label{sec:subcorpus}
Corpus size drives both retrieval difficulty and evaluation cost. A single full-corpus pass with \textsc{pplx-embed-v1-4b} takes 4{,}608 H200 GPU-hours. Even the smallest retriever we evaluate, \textsc{EmbeddingGemma-300M}~\citep{vera2025embeddinggemmapowerfullightweighttext} at 300M parameters, requires almost 200 H200 GPU-hours. Repeated, community-scale evaluation at this corpus size is therefore impractical rather than merely expensive~\citep{Scells2022GreenIR}. To construct a reliable subsampled version of the corpus, sampling must retain all relevant documents and challenging distractors. In this work, we consider a document as a challenging distractor if at least one retriever ranks it above a relevant document in its top-$k$ results. We argue that omitting such distractors can inflate evaluation scores and change the relative ranking of systems~\citep{Dinzinger2026BridgingSubsampled}. Subcorpus sampling~\citep{Froebe2025CorpusSubsampling} achieves this by carefully selecting a subset that preserves the documents needed to reproduce the original evaluation results.
\begin{figure}[t]
\centering
\small
\begin{tcolorbox}[colback=gray!5, colframe=gray!40, boxrule=0.4pt,
                  left=4pt, right=4pt, top=3pt, bottom=3pt]
\textbf{System.} You are an EXTREMELY STRICT search-relevance judge. Answer
YES only if this document satisfies EVERY specific constraint in the query ---
the exact entity, the exact time/date, the exact quantity, and the exact
aspect being asked about --- and contains the precise information the user
wants. Answer NO if any constraint is missing, mismatched, or only
approximately met (e.g.\ a similar-but-different entity, a different date, a
near but not equal value, or only part of the question). When in doubt, answer
NO. Answer YES or NO. No other text.

\vspace{6pt}\hrule\vspace{6pt}

\textbf{User.} Query: ``\texttt{<query>}''\\
Document: ``\texttt{<document>}''\\
Is this document relevant?
\end{tcolorbox}
\caption{Relevance-judging prompt for \textsc{deepseek-ai/DeepSeek-V4-Flash}. The system message enforces strict constraint matching; the user message supplies the query and the truncated document.}
\label{fig:prompt}
\end{figure}

Let $\mathcal{C}$ denote the full corpus, $P \subset \mathcal{C}$ the set of positively judged documents, and $\mathcal{Q}$ the set of evaluation queries. Let $\mathcal{M}$ denote the set of retrieval systems whose rankings are used to construct the top-$k$ pool, with $k \in \{100, 500, 1000, 2000, 3000\}$. For each retriever $m \in \mathcal{M}$ and query $q \in \mathcal{Q}$, let $R^m_{q}$ denote the ranking produced by $m$ for $q$. We formally define the top-$k$ pool as follows:
\[
U_k =
\bigcup_{q \in \mathcal{Q}}
\bigcup_{m \in \mathcal{M}}
\operatorname{top}_k(R^m_{q}).
\]
When subsampling, each sampling strategy constructs a distractor set $S \subset \mathcal{C} \setminus P$ and a subcorpus $\widehat{\mathcal{C}} = P \cup S$.
Thus, every positively judged document is retained in every subcorpus, while subsampling removes only unjudged documents. \citet{Froebe2025CorpusSubsampling} show that distractors selected from the union of highly ranked results produced by multiple retrievers can serve as hard negatives. Intuitively, these are unjudged documents that retrieval systems deem similar to a query and are therefore more challenging than uniformly sampled distractors. In this work, we compare three strategies for constructing the distractor set $S$. Because different strategies produce subcorpora of different sizes, we express size as the percentage of the full corpus.

\emph{Depth-$k$ pooling}~\citep{Froebe2025CorpusSubsampling} includes every
unjudged document that appears in the top-$k$ results of at least one retriever for at least one query, i.e., $S = U_k \setminus P$. If a document appears in the top-$k$ results of multiple retrieval systems, it occupies only one place in the subcorpus; we retain the set of unique documents.

\emph{Reciprocal rank fusion} (RRF)~\citep{Cormack2009RRF} combines the retrievers in $\mathcal{M}$ into a single ranking for each query, using a rank offset of $60$. We then add the top $k$ unjudged documents from each fused ranking that are not already in the subcorpus.

Finally, with \emph{uniform random} sampling, we sample $k$ unjudged documents uniformly for each query from those not already included in the subcorpus. Note that the query determines only how many documents are sampled, not which documents are sampled in this case. Repeating this procedure for each query produces the same subsampled corpus size as RRF.

\section{Benchmarking First-Stage Retrievers}
\label{sec:benchmark-retrievers}
We use \ours{} to benchmark thirteen retrievers, including lexical, dense, and late-interaction embedding models over $\sim$190M documents and $\sim$70k queries.
Each retriever is scored using three complementary metrics on both the full corpus and the sampled subcorpus.

\subsection{Experimental Setup}
\label{sec:experimental-setup}

\autoref{tab:models} lists the retrievers we evaluate.
All neural retrievers are open-weight and released on or after 1~January 2025 to reflect the current state of the field and do not overlap with models used to create the benchmark.
For lexical retrieval, we use BM25-tantivy as a standard lexical reference point.
For dense retrieval, we select ten open-weight multilingual encoders whose parameter counts span 0.3B to 8B, spanning six families: Qwen3-Embedding, EmbeddingGemma, Voyage~4, Nemotron-3-Embed, DenseOn, and pplx-embed-v1.
This range lets us observe how retrieval quality scales with model size.
For late-interaction retrieval, we include mLateOn and pplx-embed-v1-late-0.6B, the two open-weight multilingual late-interaction checkpoints released after January~2025 available at the time of writing.
We do not benchmark learned sparse retrievers such as SPLADE-v2~\citep{Formal2021SPLADEVS} because multilingual checkpoints for this family remain scarce.
Most IR benchmarks treat nDCG@10 as their primary metric because they target the final ranked list presented to users. \ours{} instead targets the first stage of a multi-stage ranking pipeline, whose role is to recall as many relevant documents as possible at scale. We therefore adopt Recall@1000 as our primary metric, and additionally report Recall@100 and nDCG@10 to characterize retrieval quality at shallower depths.
All three metrics are reported separately on the three label sets from \autoref{sec:qrels}.

For BM25, we use tantivy, which provides scalable inverted-index construction over the corpus and language-aware analyzers for multilingual retrieval. This is the same BM25 used for subcorpus sampling (\autoref{sec:subcorpus}).
For every dense retriever, we adopt the model's default retrieval instruction template (when available) released with its checkpoint to preserve out-of-the-box behavior.
For the two late-interaction retrievers, mLateOn and pplx-embed-v1-late-0.6B, we use 32 tokens for queries.
All retrievers encode the first 512 tokens of each document, a truncation forced by the cost of encoding a 190M-document corpus; since the mean document is roughly 3.3k tokens, most of the corpus body is never seen by any system. Spot checks on the subsampled corpus with longer inputs (1024 tokens for dense, 2048 for late-interaction) did not reorder the models, but we do not claim the 512-token scores are unbiased estimates of full-document effectiveness.

\subsection{Comparison of Sampling Strategies}
\label{sec:sampling-comparison}
In \ours{}, each of the $\sim$70k queries is associated with multiple relevant documents, making the union of per-query candidate sets substantially larger than the collection studied by \citet{Froebe2025CorpusSubsampling}. Under a fixed sampling budget, this creates a trade-off between retaining evaluation-critical documents and reducing the corpus size. We therefore investigate which sampling strategy and subcorpus size support the same conclusions in \ours{}.

Subcorpus sampling inherits the bias induced by the retrieval systems used to construct the subcorpus: a collection can favor contributing retrievers and be biased against those that did not contribute~\citep{Zobel1998Reusable,buckley2007bias}. We use a temporal holdout to evaluate subcorpus generalization to models that did not contribute to its construction. Specifically, only models released before 1~January 2025 contribute to subcorpus construction, and $\mathcal{M}$ spans lexical, dense, and late-interaction retrievers~(\autoref{app:pooling-models}), so the held-out models are tested against a pool built from all three families rather than from a single one. Every neural retriever in \autoref{tab:models} was released after that date and is held out; BM25 also contributes to the judging and sampling pools. The same nine retrievers form the judging pool of \autoref{sec:qrels}, so the holdout covers label generation as well as sampling.

\begin{figure}
     \centering
     \includegraphics[width=0.5\linewidth]{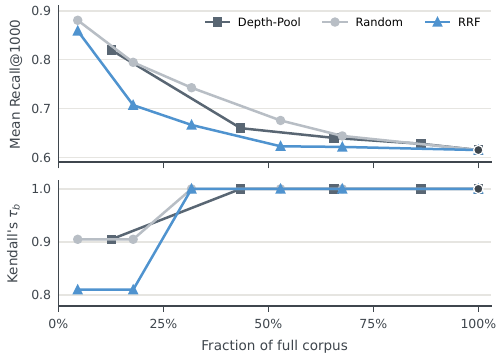}
     \caption{Mean Recall@1000 (top) and rank agreement with the full corpus (bottom, Kendall's $\tau_b$) versus the retained fraction of the corpus. RRF preserves the model ranking from $\sim30\%$ of the full corpus.}
     \label{fig:recall_subcorpus}
\end{figure}

To assess how closely each sampling strategy matches full-corpus evaluation, we measure the absolute difference in Recall@1000 and ranking agreement using Kendall's~$\tau_b$~\citep{kendall1945treatment}. 
For $\mathcal{M}$ we use the same 9 retrievers as in \autoref{sec:qrels} to avoid bias in evaluation. The positive set $P$ is taken from the \textsc{Combined+LLM-Judged} relevance set. To create the top-$k$ pool, we choose a value $k$ from the set $\{100, 500, 1000, 2000, 3000\}$ and show the comparisons in \autoref{fig:recall_subcorpus}.  RRF and uniform random both preserve the full-corpus ranking ($\tau_b = 1.00$) from $31.7\%$ of the corpus, and depth-$k$ pooling from $43.4\%$. At that budget, RRF stays closest to the full-corpus scores, inflating mean Recall@1000 by $5.1$ points against $12.8$ points for uniform random, while depth-$k$ pooling reaches comparable fidelity ($4.5$ points) only on a larger subcorpus. We therefore adopt RRF with $k=1000$ as the best trade-off between ranking fidelity, absolute error, and evaluation cost: it is the smallest RRF pool that reproduces the full-corpus ordering, and doubling it to $k=2000$ narrows the recall gap to $0.8$ points but requires $52.9\%$ of the corpus. The same \textsc{pplx-embed-v1-4b} evaluation drops to roughly 1{,}500 H200 GPU-hours, and \textsc{EmbeddingGemma-300M} to under 70. Subcorpus sampling enables faithful web-scale evaluation in about eight hours on a single 8$\times$H200 node rather than on a multi-node cluster.
We report the results in the following sections, accordingly.

\begin{table}[t!]
\caption{Retrievers evaluated in \ours{}; every neural checkpoint is
open-weight and released on or after 1~January 2025.}
\label{tab:models}
\centering
\small
\setlength{\tabcolsep}{3.5pt}
\begin{tabular}{@{}l l r r c r c@{}}
\toprule
\textbf{Model} & \textbf{Family} &
  \textbf{\#P} & \textbf{$d$} & \textbf{Vec.} &
  \textbf{Ctx} & \textbf{Rel.} \\
  \midrule
\href{https://github.com/quickwit-oss/tantivy}{BM25-tantivy} \citep{tantivy}
  & lexical & --- & --- & single & --- & --- \\
\midrule
\href{https://huggingface.co/google/embeddinggemma-300m}{EmbeddingGemma-300M}
  \citep{vera2025embeddinggemmapowerfullightweighttext}
  & dense & 0.3B & 768 & single & 2k & 2025-09 \\
\href{https://huggingface.co/lightonai/mDenseOn}{mDenseOn}
  \citep{sourty2026denseonlateonfullyopen}
  & dense & 0.3B & 768 & single & 8k & 2026-07 \\
\href{https://huggingface.co/nvidia/Nemotron-3-Embed-1B-BF16}{Nemotron-3-Embed-1B}
\citep{babakhin2026nemotron3embed}
  & dense & 1.1B & 2048 & single & 32k & 2026-07 \\
\href{https://huggingface.co/nvidia/Nemotron-3-Embed-8B-BF16}{Nemotron-3-Embed-8B}
  & dense & 8B & 4096 & single & 32k & 2026-07 \\
\href{https://huggingface.co/perplexity-ai/pplx-embed-v1-0.6b}{pplx-embed-v1-0.6b}
  \citep{eslami-etal-2026-diffusion}
  & dense & 0.6B & 1024 & single & 32k & 2026-01 \\
\href{https://huggingface.co/perplexity-ai/pplx-embed-v1-4b}{pplx-embed-v1-4b}
  \citep{eslami-etal-2026-diffusion}
  & dense & 4B & 2560 & single & 32k & 2026-01 \\
\href{https://huggingface.co/Qwen/Qwen3-Embedding-0.6B}{Qwen3-Embedding-0.6B}
  \citep{zhang2025qwen3embeddingadvancingtext}
  & dense & 0.6B & 1024 & single & 32k & 2025-06 \\
\href{https://huggingface.co/Qwen/Qwen3-Embedding-4B}{Qwen3-Embedding-4B}
  & dense & 4B & 2560 & single & 32k & 2025-06 \\
\href{https://huggingface.co/Qwen/Qwen3-Embedding-8B}{Qwen3-Embedding-8B}
  & dense & 8B & 4096 & single & 32k & 2025-06 \\
\href{https://huggingface.co/voyageai/voyage-4-nano}{voyage-4-nano}
  & dense & 0.3B & 2048 & single & 32k & 2026-01 \\
\midrule
\href{https://huggingface.co/lightonai/mLateOn}{mLateOn}
  \citep{sourty2026denseonlateonfullyopen}
  & late & 0.3B & 128 & multi & 8k & 2026-07 \\
\href{https://huggingface.co/perplexity-ai/pplx-embed-v1-late-0.6b}{pplx-embed-v1-late-0.6B}~\citep{eslami-etal-2026-diffusion}
  & late & 0.6B & 128 & multi & 32k & 2026-03 \\
\bottomrule
\end{tabular}
\end{table}

\subsection{Results}
\label{sec:main_results}

\paragraph{Main.} \autoref{tab:qrel-results-combined} reports Recall@1000, Recall@100, and nDCG@10 on the three relevance-judgment sets, evaluated on both the full 190M-document corpus and the RRF-subsampled corpus. For simplicity, we refer to the \textsc{Combined+LLM-judged} relevance set as Comb. in this section.
On the primary metric Recall@1000, the winner in each of the three judgment sets is identical between the full and subsampled evaluations: pplx-embed-v1-4b wins all three (Citation, Web Ranking, and Comb.); subsampling raises every Recall@1000 score by 2--7 points but leaves the winner on each set unchanged.
Among sub-1B dense encoders, pplx-embed-v1-0.6b attains the strongest Recall@1000 across all three passes, followed by EmbeddingGemma-300M, with mDenseOn, Qwen3-Embedding-0.6B, and voyage-4-nano trailing EmbeddingGemma-300M by 6--9 points on Comb.
Within the Qwen3-Embedding family, Recall@1000 improves monotonically with model size from 0.6B to 4B to 8B across all three qrel passes ($58.68 \to 62.78 \to 65.32$ on Comb.).
Comparing late interaction models with dense encoders yields mixed results. mLateOn outperforms mDenseOn, whereas pplx-embed-v1-late-0.6B trails pplx-embed-v1-0.6b. Both late-interaction models fall below the strongest dense encoders at their respective sizes on Recall@1000, by 3--5 points on Comb., but the gaps narrow to at most 1 point on nDCG@10, where mLateOn also outperforms Qwen3-Embedding-4B.
The leader depends on the metric: pplx-embed-v1-4b wins our primary metric Recall@1000, but Nemotron-3-Embed-8B is strongest on both Recall@100 and nDCG@10 under Combined, placing more of its relevant documents near the top of the ranking while recovering fewer overall.

\begin{figure*}[t]
\vspace{-1mm}
    \centering
    \includegraphics[width=0.95\textwidth]{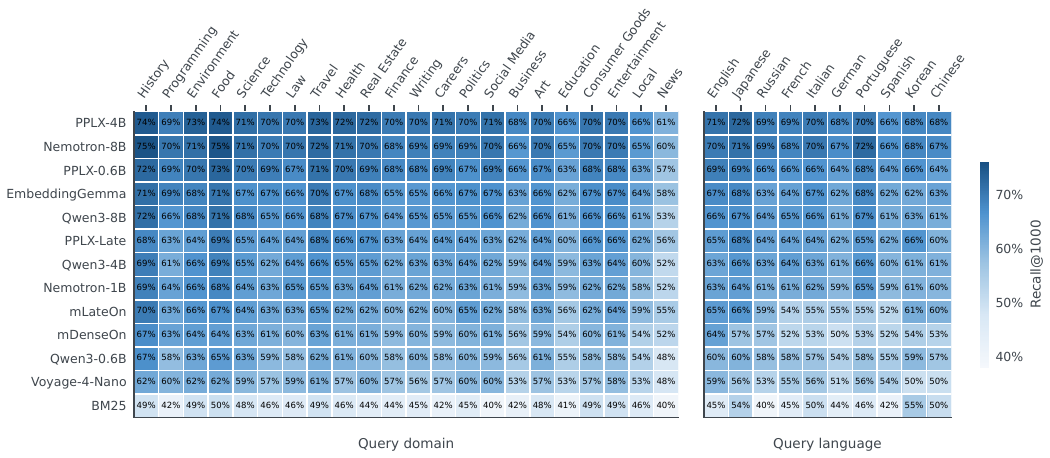}
    \vspace{-2mm}
    \caption{Recall@1000 under Combined+LLM-Judged relevance, broken down by query domain (left) and query language (right). Rows represent retrievers and columns represent query slices; cell labels report recall as percentages. Both panels use the same color scale, with darker cells indicating higher recall.}
    \vspace{-1mm}
    \label{fig:slice}
\end{figure*}

\paragraph{Domain-specific Results.} In \autoref{fig:slice}, we present the performance of each embedding model across domains. For simplicity, we only report results using the \textsc{Combined+LLM-Judged} judgments. Our results show that pplx-embed-v1-4b performs strongly across most domains, often matching or
outperforming Nemotron-3-Embed-8B, and pplx-embed-v1-0.6b ranks third with results similar to EmbeddingGemma-300M. BM25-tantivy achieves the lowest recall in every domain.

\paragraph{Language-specific Results.}
\autoref{fig:slice} reports Recall@1000 by query language. pplx-embed-v1-4b and Nemotron-3-Embed-8B are the strongest models overall. Nemotron-3-Embed-8B leads pplx-embed-v1-4b by about 2 percentage points on Portuguese and by less than a point on Spanish, Russian, Italian, and Korean, while matching or trailing it on the remaining five languages. pplx-embed-v1-0.6b ranks third overall, followed by EmbeddingGemma-300M. mDenseOn scores 64\% on English queries and 49--57\% on queries in other languages. mLateOn scores 65\% on English queries and 52--66\% elsewhere, with Japanese (66\%) slightly exceeding English. Thus, both models perform worse in most non-English languages, although the size of the gap varies by language.

\begin{figure}
    \centering
    \includegraphics[width=0.5\linewidth]{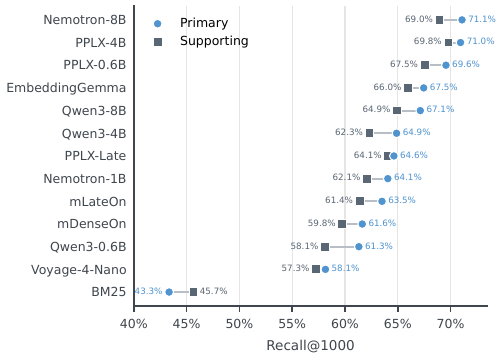}
    \vspace{-1mm}
    \caption{Recall@1000 for primary and supporting agent-reformulated queries. Each line connects the two scores for one retriever. Neural retrievers perform worse on supporting queries, while BM25 improves slightly.}
    \vspace{-2mm}
    \label{fig:primary-vs-supporting}
\end{figure}

\paragraph{Query-type Results.} \autoref{fig:primary-vs-supporting} reports Recall@1000 separately for the primary and supporting reformulated queries. All neural retrievers perform substantially better on primary queries and are markedly weaker on supporting queries. This reflects the different roles of the two query types: primary reformulated queries stay close to the user’s original information need, while supporting reformulated queries are explicitly generated by the agent to collect additional, more specific information. The performance drop on supporting queries may also stem from how these embedding models are usually trained, i.e., most embedding models are trained predominantly on human-written queries, which resemble primary queries more than agent-generated supporting queries, and thus may be less effective at representing the latter.

BM25, however, shows the opposite trend, with slightly higher performance on supporting queries, but the gap is only about 2.3 percentage points and its overall performance remains much lower than that of neural retrievers.

\begin{table*}[th!]\centering
\vspace{-1mm}
\caption{Retrieval results across relevance sets, metrics, and evaluation corpora. Each cell reports two numbers: \textit{Full} on the 190M-document corpus and \textit{Sub} on the RRF-subsampled corpus ($k{=}1000$). Best value in bold for Recall@1000, our primary metric. Comb. denotes \textsc{Combined+LLM-judged} case.}
\vspace{-1mm}
\label{tab:qrel-results-combined}\small\setlength{\tabcolsep}{2.5pt}\resizebox{\textwidth}{!}{%
\begin{tabular}{@{}l cc cc cc cc cc cc cc cc cc cc@{}}
\toprule
& \multicolumn{6}{c}{Recall@1000} & \multicolumn{6}{c}{Recall@100} & \multicolumn{6}{c}{nDCG@10} \\
\cmidrule(lr){2-7} \cmidrule(lr){8-13} \cmidrule(lr){14-19}
& \multicolumn{2}{c}{Citation} & \multicolumn{2}{c}{Web Ranking} & \multicolumn{2}{c}{Comb.}
& \multicolumn{2}{c}{Citation} & \multicolumn{2}{c}{Web Ranking} & \multicolumn{2}{c}{Comb.}
& \multicolumn{2}{c}{Citation} & \multicolumn{2}{c}{Web Ranking} & \multicolumn{2}{c}{Comb.} \\
\cmidrule(lr){2-3} \cmidrule(lr){4-5} \cmidrule(lr){6-7}
\cmidrule(lr){8-9} \cmidrule(lr){10-11} \cmidrule(lr){12-13}
\cmidrule(lr){14-15} \cmidrule(lr){16-17} \cmidrule(lr){18-19}
Model & F & S & F & S & F & S & F & S & F & S & F & S & F & S & F & S & F & S \\
\midrule
BM25-tantivy             & 60.84 & 65.58 & 42.50 & 47.61 & 45.25 & 50.38 & 32.80 & 32.91 & 16.97 & 17.07 & 18.33 & 18.44 & 7.30 & 7.30 & 11.45 & 11.46 & 29.27 & 29.30 \\
\midrule
EmbeddingGemma-300M      & 80.10 & 83.25 & 58.93 & 63.75 & 66.24 & 70.93 & 49.69 & 50.34 & 25.02 & 25.54 & 27.70 & 28.30 & 12.21 & 12.25 & 17.30 & 17.37 & 41.95 & 42.11 \\
mDenseOn                 & 73.72 & 77.37 & 52.56 & 57.41 & 60.09 & 64.95 & 43.17 & 43.93 & 21.70 & 22.28 & 24.20 & 24.87 & 10.08 & 10.13 & 15.06 & 15.17 & 38.87 & 39.09 \\
Nemotron-3-Embed-1B      & 75.64 & 79.64 & 54.37 & 59.72 & 62.44 & 67.69 & 43.15 & 44.17 & 21.76 & 22.49 & 25.15 & 26.02 & 9.69 & 9.75 & 14.83 & 14.95 & 38.84 & 39.20 \\
Nemotron-3-Embed-8B      & 83.14 & 85.78 & 61.73 & 66.29 & 69.33 & 73.56 & 53.33 & 53.93 & 26.88 & 27.41 & 30.32 & 30.93 & 13.17 & 13.21 & 18.69 & 18.76 & 45.65 & 45.80 \\
pplx-embed-v1-0.6b       & 81.85 & 84.66 & 62.15 & 66.70 & 67.91 & 72.46 & 52.14 & 52.69 & 27.81 & 28.28 & 28.62 & 29.13 & 12.52 & 12.54 & 19.76 & 19.82 & 42.73 & 42.85 \\
pplx-embed-v1-4b         & \textbf{85.03} & \textbf{87.60} & \textbf{65.73} & \textbf{70.63} & \textbf{70.00} & \textbf{75.01} & 56.54 & 57.32 & 30.28 & 30.95 & 30.17 & 30.87 & 14.18 & 14.22 & 21.29 & 21.40 & 44.12 & 44.34 \\
Qwen3-Embedding-0.6B     & 71.19 & 75.56 & 51.82 & 57.38 & 58.68 & 64.37 & 39.42 & 40.14 & 20.63 & 21.21 & 22.75 & 23.43 & 8.53 & 8.57 & 14.11 & 14.19 & 36.19 & 36.40 \\
Qwen3-Embedding-4B       & 76.49 & 80.45 & 55.60 & 61.35 & 62.78 & 68.41 & 43.96 & 44.83 & 22.33 & 22.99 & 25.19 & 25.97 & 9.91 & 9.95 & 14.97 & 15.06 & 38.82 & 39.09 \\
Qwen3-Embedding-8B       & 79.26 & 82.69 & 58.34 & 63.64 & 65.32 & 70.43 & 47.30 & 48.09 & 24.14 & 24.73 & 26.95 & 27.63 & 10.75 & 10.80 & 16.27 & 16.36 & 41.20 & 41.39 \\
voyage-4-nano            & 69.55 & 75.18 & 50.14 & 57.03 & 57.42 & 64.37 & 36.37 & 37.89 & 18.92 & 19.99 & 21.43 & 22.69 & 6.96 & 7.06 & 11.77 & 12.01 & 30.59 & 31.38 \\
\midrule
mLateOn                  & 74.16 & 77.71 & 53.12 & 57.92 & 61.77 & 66.46 & 44.76 & 45.55 & 22.55 & 23.21 & 26.32 & 27.10 & 10.94 & 11.03 & 15.51 & 15.68 & 41.61 & 41.98 \\
pplx-embed-v1-late-0.6B  & 78.64 & 82.43 & 58.14 & 63.80 & 64.20 & 69.80 & 48.44 & 49.38 & 25.01 & 25.77 & 27.34 & 28.21 & 11.88 & 11.95 & 17.25 & 17.39 & 41.77 & 42.14 \\
\bottomrule
\end{tabular}}
\end{table*}

\section{Insights and Analysis}
\label{sec:insights}        
We restrict the analysis in this section to the \textsc{Combined+LLM-Judged} label set and Recall@1000 rather than repeating it for every label set and metric. \autoref{sec:main_results} shows that retriever rankings are largely stable across the three label sets.

\subsection{Unique Positive Coverage}
\label{sec:unique-positives}      
Which positives a retriever recovers is not ordered by its aggregate recall. BM25-tantivy reaches the lowest Recall@1000 of any retriever ($45.25$) yet contributes the most unique positives, $36{,}073$, about $1.3$ times the next-highest contributor. That runner-up is pplx-embed-v1-4b with $27{,}979$, which also attains the highest Recall@1000 ($70.00$), followed by EmbeddingGemma-300M with $17{,}033$, which ranks fourth of thirteen ($66.24$) at 300M parameters. The remaining retrievers contribute fewer unique positives: pplx-late-0.6b ($11{,}556$) and pplx-embed-v1-0.6b ($11{,}430$), then Nemotron-3-Embed-8B ($10{,}506$, second on Recall@1000 at $69.33$) and mLateOn ($8{,}536$), with Nemotron-3-Embed-1B contributing the fewest at $2{,}938$. Retrievers of different families and training recipes therefore recover different parts of the relevant set, and the retriever with the weakest aggregate recall still recovers the largest set that no other retriever recovers.

BM25 may therefore still be useful alongside stronger neural retrievers. Comparing such combinations with ensembles containing only neural retrievers, using the same number of retrievers, fusion rule, and candidate budget, would test whether its unique positives lead to higher recall.

\subsection{Hard Negatives and Hard Positives}
We complement the aggregate results with an error analysis of the hardest cases in both directions: documents that all thirteen retrievers rank in their top-1000 despite being labeled non-relevant, and relevant documents that none of the thirteen ranks in its top-1000. Both taxonomies are defined over the \textsc{Combined+LLM-Judged} label set. We classify both sets with \textsc{deepseek-ai/DeepSeek-V4-Flash}, the same model that produced the LLM-generated part of that label set, so the hard-negative analysis asks the judge to revisit its own decisions. We classify a uniform random sample of $4{,}000$ hard negatives; the hard-positive set has $298{,}868$ members ($6.3\%$ of all relevant documents), from which we likewise classify a uniform random sample of $4{,}000$. The two prompts are shown in \autoref{fig:prompts}.

\paragraph{Hard Negatives.}
Our review finds that $16.8\%$ of the sampled hard negatives are in fact relevant. We treat these as label false negatives rather than excluding them from the pool, since our relevance pool is built only from models released before 2025 and, of the thirteen evaluated retrievers, only BM25 contributed to the pool. Their absence from the relevance scores reflects a gap in the pooling models rather than a bias toward or against any system under evaluation. The remaining $83.2\%$ are genuine hard negatives. Of all sampled pairs, the dominant relationships are: \textsc{wrong-aspect}, where the document shares the query's general topic but addresses a different aspect or subtopic ($35.6\%$); \textsc{partial}, where the document is on-topic and touches the need but is incomplete or only tangential ($19.6\%$); \textsc{related-entity}, where the document concerns a similar but distinct product, person, place, or version ($14.5\%$); \textsc{generic}, where boilerplate, navigational, or listing pages match the query's keywords without substantive content ($7.5\%$); and \textsc{temporal}, where the document is the right topic and entity but from the wrong point in time ($4.8\%$). The remaining $1.2\%$ fall outside the taxonomy.
\begin{figure}[t]
    \centering
    \small
    \begin{tcolorbox}[colback=gray!5, colframe=gray!40, boxrule=0.4pt,
                      left=4pt, right=4pt, top=3pt, bottom=3pt]
    \textbf{Hard negatives.} You are given a query and a document that are ranked top-1000 but that the labels mark NON-relevant. Choose
    EXACTLY ONE label: \textsc{false\_negative} (the document is actually
    relevant; the label is wrong), \textsc{wrong\_aspect} (same topic but a
    different aspect/subtopic than asked), \textsc{related\_entity} (a similar
    but different product, person, place, or version), \textsc{partial} (touches
    the need but incomplete or only tangential), \textsc{generic}
    (boilerplate/navigational/listing page matching keywords, no substance), or
    \textsc{temporal} (right topic but wrong time period / outdated / different
    edition). Otherwise answer \textsc{other}. End with: LABEL: \texttt{<one
    label>}
    
    \vspace{6pt}\hrule\vspace{6pt}
    
    \textbf{Hard positives.} You are given a query and a document that IS
    relevant but which the retriever FAILED to rank in its top results. Choose
    the single most likely reason: \textsc{lexical} (relevant by meaning but
    shares almost none of the query's words), \textsc{entity\_numeric} (relevance
    hinges on an exact entity, code, quantity, or date), \textsc{truncation}
    (query-relevant content sits deep inside a long document),
    \textsc{multilingual} (query and document in different languages or
    scripts), \textsc{reasoning} (linking query to document needs multi-step
    inference / world knowledge), or \textsc{generic\_dup} (generic/boilerplate
    or one of many near-duplicates). Otherwise answer \textsc{other}. End with:
    LABEL: \texttt{<one label>}
    \end{tcolorbox}
    \vspace{-2mm}
    \caption{Hard-negative and hard-positive taxonomy prompts for \textsc{deepseek-ai/DeepSeek-V4-Flash}. Each prompt gives the query as text and the document as its title and body. 
    }
    \vspace{-2mm}
    \label{fig:prompts}
\end{figure}

\paragraph{Hard Positives.}
Across a $4{,}000$-document sample of the $298{,}868$ relevant documents that none of the thirteen retrievers ranks in its top-1000, \textsc{lexical}, where the document is relevant by meaning but shares almost none of the query's words or phrasing, is the largest category ($28.3\%$). \textsc{reasoning}, where linking the query to the document requires multi-step inference or world knowledge, and \textsc{truncation}, where the query-relevant content sits deep inside a long document, account for a further $21.7\%$ and $18.2\%$, respectively. The remaining categories are smaller: \textsc{generic\_dup}, generic or boilerplate content or one of many near-duplicates ($9.7\%$); \textsc{multilingual}, where query and document are written in different languages or scripts ($9.4\%$); \textsc{entity\_numeric}, where relevance hinges on a specific named entity, code, quantity, or date ($4.3\%$); and $8.4\%$ outside the taxonomy.

\section{Conclusion}
\label{sec:conclusion}
We introduce \ours{}, a retrieval benchmark built from nine months of production search traffic, with \(\sim\)190M documents, \(\sim\)70k agent-reformulated queries, and three relevance-label sets. Across thirteen lexical, dense, and late-interaction retrievers, rankings are largely stable across label sets, while the retriever with the lowest aggregate recall contributes the most unique labeled positives. A \(\sim\)30\% RRF subcorpus preserves the ranking order under the Combined judgments, with a small, systematic inflation of absolute recall. For a 4B embedding model, this reduces evaluation cost from 4{,}608 to roughly 1{,}500 H200 GPU-hours.

The many labeled positives per query let us measure which relevant documents each retriever adds to an ensemble. Future work can use this to test whether diverse ensembles recover more relevant documents than equally sized ensembles of stronger but more similar retrievers. We release a \href{https://huggingface.co/spaces/perplexity-ai/q2d-web-leaderboard}{public leaderboard} for continued evaluation on \ours{}, while keeping the queries and relevance judgments private to limit training contamination.

\section*{Ethics and Privacy Statement}

\ours{} is derived from production search traffic of a commercial conversational search assistant, and we confirm that use of this traffic is consistent with the applicable user agreements, privacy policy, and data protection law, and apply PII detection during query sampling to exclude any query flagged as containing PII. \ours{} remains a private benchmark rather than a released dataset, with only aggregate results such as leaderboard scores shared externally, and we maintain its three relevance-judgment sources separately rather than treating any single pass as ground truth. We do not foresee a use of \ours{} beyond the standard dual-use profile of retrieval research, in which a stronger retriever surfaces both beneficial and harmful information equally well.

\bibliography{main}

@inproceedings{Craswell2023TRECDL23,
  author = {Nick Craswell and Bhaskar Mitra and Emine Yilmaz and Hossein A. Rahmani and Daniel Campos and Jimmy Lin and Ellen M. Voorhees and Ian Soboroff},
  booktitle = {TREC},
  doi = {10.6028/nist.sp.1266.deep-overview},
  publisher = {National Institute of Standards and Technology (NIST)},
  title = {Overview of the TREC 2023 Deep Learning Track.},
  url = {https://trec.nist.gov/pubs/trec32/papers/Overview_deep.pdf},
  year = {2023}
}

@article{Chen2024MSWS,
  author = {Qi Chen and Xiubo Geng and Corby Rosset and Carolyn Buractaon and Jingwen Lu and Tao Shen and Kun Zhou and Chenyan Xiong and Yeyun Gong and Paul N. Bennett and Nick Craswell and Xing Xie and Fan Yang and Bryan Tower and Nikhil Rao and Anlei Dong and Wenqi Jiang and Zheng Liu and Mingqin Li and Chuanjie Liu and Zengzhong Li and Rangan Majumder and Jennifer Neville and Andy Oakley and K. Risvik and H. Simhadri and Manik Varma and Yujing Wang and Linjun Yang and Mao Yang and Ce Zhang},
  doi = {10.1145/3589335.3648327},
  journal = {The Web Conference},
  pages = {292-301},
  title = {MS MARCO Web Search: A Large-scale Information-rich Web Dataset with Millions of Real Click Labels},
  url = {https://doi.org/10.1145/3589335.3648327},
  year = {2024},
  
}

@article{Bajaj2016MSMARCO,
  archiveprefix = {arXiv},
  author = {Payal Bajaj and Daniel Campos and Nick Craswell and Li Deng and Jianfeng Gao and
Xiaodong Liu and Rangan Majumder and Andrew McNamara and Bhaskar Mitra and
Tri Nguyen and Mir Rosenberg and Xia Song and Alina Stoica and Saurabh Tiwary and
Tong Wang},
  doi = {10.48550/arXiv.1611.09268},
  eprint = {1611.09268},
  journal = {arXiv preprint arXiv:1611.09268},
  primaryclass = {cs.CL},
  title = {{MS MARCO}: A Human Generated MAchine Reading COmprehension Dataset},
  url = {https://arxiv.org/abs/1611.09268},
  year = {2016}
}

@inproceedings{Chen2025AIRBench,
  author = {Jianlyu Chen and Nan Wang and Chaofan Li and Bo Wang and Shitao Xiao and Han Xiao and Hao Liao and Defu Lian and Zheng Liu},
  booktitle = {ACL},
  doi = {10.18653/v1/2025.acl-long.982},
  pages = {19991-20022},
  publisher = {Association for Computational Linguistics},
  title = {AIR-Bench: Automated Heterogeneous Information Retrieval Benchmark.},
  url = {https://doi.org/10.18653/v1/2025.acl-long.982},
  year = {2025}
}

@article{Pradeep2024Ragnarok,
      title={Ragnar\"ok: A Reusable RAG Framework and Baselines for TREC 2024 Retrieval-Augmented Generation Track}, 
      author={Ronak Pradeep and Nandan Thakur and Sahel Sharifymoghaddam and Eric Zhang and Ryan Nguyen and Daniel Campos and Nick Craswell and Jimmy Lin},
      year={2024},
      eprint={2406.16828},
      archivePrefix={arXiv},
      primaryClass={cs.IR},
      url={https://arxiv.org/abs/2406.16828}, 
      journal = {arXiv preprint}
}

@article{Upadhyay2024LargeScaleLLMAssess,
  archiveprefix = {arXiv},
  author = {Shivani Upadhyay and Ronak Pradeep and Nandan Thakur and Daniel Campos and
Nick Craswell and Ian Soboroff and Hoa Trang Dang and Jimmy Lin},
  doi = {10.1145/3731120.3744605},
  eprint = {2411.08275},
  journal = {arXiv preprint arXiv:2411.08275},
  pages = {358-368},
  primaryclass = {cs.IR},
  title = {A Large-Scale Study of Relevance Assessments with Large Language Models:
An Initial Look},
  url = {https://arxiv.org/abs/2411.08275},
  year = {2024}
}

@misc{ClueWeb12,
  author = {{The Lemur Project}},
  howpublished = {\url{https://lemurproject.org/clueweb12/}},
  note = {733{,}019{,}372 English web pages crawled February--May 2012;
the ClueWeb12-B13 subset contains 52{,}343{,}021 documents},
  title = {The {ClueWeb12} Dataset},
  year = {2012}
}

@misc{ClueWeb09,
  author = {{The Lemur Project}},
  howpublished = {\url{https://lemurproject.org/clueweb09/}},
  note = {1{,}040{,}809{,}705 web pages crawled January--February 2009;
Category B is the first 50M English pages},
  title = {The {ClueWeb09} Dataset},
  year = {2009}
}

@inproceedings{Froebe2025CorpusSubsampling,
  author = {Maik Fr{\"o}be and Andrew Parry and Harrisen Scells and Shuai Wang and
Shengyao Zhuang and Guido Zuccon and Martin Potthast and Matthias Hagen},
  booktitle = {Advances in Information Retrieval -- 47th European Conference on Information
Retrieval (ECIR 2025), Proceedings, Part I},
  doi = {10.1007/978-3-031-88708-6_29},
  pages = {453--471},
  publisher = {Springer},
  series = {Lecture Notes in Computer Science},
  title = {Corpus Subsampling: Estimating the Effectiveness of Neural Retrieval Models
on Large Corpora},
  url = {https://downloads.webis.de/publications/papers/froebe_2025c.pdf},
  volume = {15572},
  year = {2025}
}

@inproceedings{Thakur2021BEIR,
  archiveprefix = {arXiv},
  author = {Nandan Thakur and Nils Reimers and Andreas R{\"u}ckl{\'e} and
Abhishek Srivastava and Iryna Gurevych},
  booktitle = {Proceedings of the Thirty-fifth Conference on Neural Information Processing
Systems Datasets and Benchmarks Track (NeurIPS 2021)},
  doi = {10.48550/arXiv.2104.08663},
  eprint = {2104.08663},
  primaryclass = {cs.IR},
  title = {{BEIR}: A Heterogenous Benchmark for Zero-shot Evaluation of Information
Retrieval Models},
  url = {https://arxiv.org/abs/2104.08663},
  year = {2021}
}

@inproceedings{Scells2022GreenIR,
  author = {Harrisen Scells and Shengyao Zhuang and Guido Zuccon},
  booktitle = {Proceedings of the 45th International ACM SIGIR Conference on Research and
Development in Information Retrieval (SIGIR '22)},
  doi = {10.1145/3477495.3531766},
  pages = {2825--2837},
  publisher = {ACM},
  title = {Reduce, Reuse, Recycle: Green Information Retrieval Research},
  url = {https://doi.org/10.1145/3477495.3531766},
  year = {2022}
}

@inproceedings{Zobel1998Reusable,
  author = {Justin Zobel},
  booktitle = {Proceedings of the 21st Annual International ACM SIGIR Conference on Research
and Development in Information Retrieval (SIGIR '98)},
  doi = {10.1145/290941.291014},
  pages = {307--314},
  publisher = {ACM},
  title = {How reliable are the results of large-scale information retrieval experiments?},
  url = {https://doi.org/10.1145/290941.291014},
  year = {1998}
}

@inproceedings{Lewis2020RAG,
  author = {Patrick S. H. Lewis and
Ethan Perez and
Aleksandra Piktus and
Fabio Petroni and
Vladimir Karpukhin and
Naman Goyal and
Heinrich K{\"{u}}ttler and
Mike Lewis and
Wen{-}tau Yih and
Tim Rockt{\"{a}}schel and
Sebastian Riedel and
Douwe Kiela},
  bibsource = {dblp computer science bibliography, https://dblp.org},
  booktitle = {Advances in Neural Information Processing Systems 33: Annual Conference
on Neural Information Processing Systems 2020, NeurIPS 2020, December
6-12, 2020, virtual},
  editor = {Hugo Larochelle and
Marc'Aurelio Ranzato and
Raia Hadsell and
Maria{-}Florina Balcan and
Hsuan{-}Tien Lin},
  title = {Retrieval-Augmented Generation for Knowledge-Intensive {NLP} Tasks},
  url = {https://proceedings.neurips.cc/paper/2020/hash/6b493230205f780e1bc26945df7481e5-Abstract.html},
  year = {2020}
}

@inproceedings{Chen2025BrowseCompPlus,
    title = "{B}rowse{C}omp-Plus: A Fair and Disentangled Evaluation Benchmark for Deep Search Agents",
    author = "Chen, Zijian  and
      Ma, Xueguang  and
      Zhuang, Shengyao  and
      Nie, Ping  and
      Zou, Kai  and
      Sharifymoghaddam, Sahel  and
      Liu, Andrew  and
      Green, Joshua  and
      Patel, Kshama  and
      Meng, Ruoxi  and
      Su, Mingyi  and
      Li, Yanxi  and
      Hong, Haoran  and
      Shi, Xinyu  and
      Liu, Xuye  and
      Oyarhoseini, Hosna  and
      Thakur, Nandan  and
      Zhang, Crystina  and
      Gao, Luyu  and
      Chen, Wenhu  and
      Lin, Jimmy",
    editor = "Liakata, Maria  and
      Moreira, Viviane P.  and
      Zhang, Jiajun  and
      Jurgens, David",
    booktitle = "Proceedings of the 64th Annual Meeting of the {A}ssociation for {C}omputational {L}inguistics (Volume 1: Long Papers)",
    month = jul,
    year = "2026",
    address = "San Diego, California, United States",
    publisher = "Association for Computational Linguistics",
    url = "https://aclanthology.org/2026.acl-long.1023/",
    doi = "10.18653/v1/2026.acl-long.1023",
    pages = "22349--22370",
    ISBN = "979-8-89176-390-6"
}

@misc{OpenAI2025DeepResearch,
  author = {{OpenAI}},
  howpublished = {\url{https://openai.com/index/introducing-deep-research/}},
  note = {Accessed 2026-07-24},
  title = {Introducing Deep Research},
  year = {2025}
}

@inproceedings{Craswell2019TRECDL19,
  archiveprefix = {arXiv},
  author = {Nick Craswell and Bhaskar Mitra and Emine Yilmaz and Daniel Campos and
Ellen M. Voorhees},
  booktitle = {Proceedings of the Twenty-Eighth Text REtrieval Conference (TREC 2019)},
  eprint = {2003.07820},
  primaryclass = {cs.IR},
  publisher = {National Institute of Standards and Technology {(NIST)}},
  series = {{NIST} Special Publication},
  title = {Overview of the {TREC} 2019 Deep Learning Track},
  url = {https://trec.nist.gov/pubs/trec28/papers/OVERVIEW.DL.pdf},
  volume = {1250},
  year = {2019}
}

@article{Overwijk2022CW22,
  archiveprefix = {arXiv},
  author = {Arnold Overwijk and Chenyan Xiong and Xiao Liu and Cameron VandenBerg and
Jamie Callan},
  doi = {10.48550/arXiv.2211.15848},
  eprint = {2211.15848},
  journal = {arXiv preprint arXiv:2211.15848},
  primaryclass = {cs.IR},
  title = {{ClueWeb22}: 10 Billion Web Documents with Visual and Semantic Information},
  url = {https://arxiv.org/abs/2211.15848},
  year = {2022}
}

@article{Xi2025DeepSearchSurvey,
  author = {Xi, Yunjia and Lin, Jianghao and Xiao, Yongzhao and Zhou, Zheli and
Shan, Rong and Gao, Te and Zhu, Jiachen and Liu, Weiwen and
Yu, Yong and Zhang, Weinan},
  journal = {ArXiv preprint},
  title = {A Survey of {LLM}-based Deep Search Agents: Paradigm, Optimization,
Evaluation, and Challenges},
  url = {https://arxiv.org/abs/2508.05668},
  volume = {abs/2508.05668},
  year = {2025}
}

@inproceedings{Dinzinger2026BridgingSubsampled,
  author = {Dinzinger, Michael and Ghosh Dastidar, Kanishka and Caspari, Laura and Mitrovi{\'c}, Jelena and Granitzer, Michael},
  booktitle = {Proceedings of the 2026 International ACM SIGIR Conference on Innovative Concepts and Theories in Information Retrieval (ICTIR '26)},
  doi = {10.1145/3805713.3820404},
  pages = {56--61},
  publisher = {ACM},
  title = {Bridging the Gap between Subsampled and Full-Corpus Evaluation},
  url = {https://doi.org/10.1145/3805713.3820404},
  year = {2026}
}

@article{Broder1997SyntacticClustering,
  author = {Broder, Andrei Z. and Glassman, Steven C. and Manasse, Mark S. and Zweig, Geoffrey},
  doi = {10.1016/s0169-7552(97)00031-7},
  journal = {Computer Networks and ISDN Systems},
  number = {8--13},
  pages = {1157--1166},
  title = {Syntactic Clustering of the Web},
  url = {https://doi.org/10.1016/s0169-7552(97)00031-7},
  volume = {29},
  year = {1997}
}

@misc{zhang2025qwen3embeddingadvancingtext,
  archiveprefix = {arXiv},
  author = {Yanzhao Zhang and Mingxin Li and Dingkun Long and Xin Zhang and Huan Lin and Baosong Yang and Pengjun Xie and An Yang and Dayiheng Liu and Junyang Lin and Fei Huang and Jingren Zhou},
  eprint = {2506.05176},
  primaryclass = {cs.CL},
  title = {Qwen3 Embedding: Advancing Text Embedding and Reranking Through Foundation Models},
  url = {https://arxiv.org/abs/2506.05176},
  year = {2025}
}

@misc{babakhin2026nemotron3embed,
  author = {Babakhin, Yauhen and Ak, Ronay and Cai, Jiarui and Raman, Vinay
and Osmulski, Radek and Zakrzewski, Jakub and Gupta, Anmol
and Holworthy, Oliver and Sharifymoghaddam, Sahel and Pham, Khang
and Rong, James and Han, Steve and Sodha, Sean
and Hulseman, Isabel and Liu, Bo},
  howpublished = {Hugging Face Blog},
  month = {July},
  note = {Accessed: 2026-08-04},
  title = {{NVIDIA} {Nemotron} 3 {Embed} Ranks \#1 Overall on {RTEB},
Advancing Agentic Retrieval},
  url = {https://huggingface.co/blog/nvidia/nemotron-3-embed-wins-rteb},
  year = {2026}
}

@misc{tantivy,
  author = {Masurel, Paul and {Quickwit, Inc.} and {Tantivy contributors}},
  license = {MIT},
  note = {GitHub repository, accessed 2026-08-04},
  title = {Tantivy: A Full-Text Search Engine Library Inspired by Apache Lucene and Written in Rust},
  howpublished = {https://github.com/quickwit-oss/tantivy},
  version = {0.26.1},
  year = {2026}
}

@misc{sourty2026denseonlateonfullyopen,
      title={DenseOn with the LateOn: Fully Open Dense and Late-Interaction Models for Multilingual, Long-Context, and Code Search}, 
      author={Raphaël Sourty and Antoine Chaffin and Paulo Roberto Moura Junior and Amélie Chatelain},
      year={2026},
      eprint={2607.27178},
      archivePrefix={arXiv},
      primaryClass={cs.CL},
      url={https://arxiv.org/abs/2607.27178}, 
}

@misc{vera2025embeddinggemmapowerfullightweighttext,
  archiveprefix = {arXiv},
  author = {Henrique Schechter Vera and Sahil Dua and Biao Zhang and Daniel Salz and Ryan Mullins and Sindhu Raghuram Panyam and Sara Smoot and Iftekhar Naim and Joe Zou and Feiyang Chen and Daniel Cer and Alice Lisak and Min Choi and Lucas Gonzalez and Omar Sanseviero and Glenn Cameron and Ian Ballantyne and Kat Black and Kaifeng Chen and Weiyi Wang and Zhe Li and Gus Martins and Jinhyuk Lee and Mark Sherwood and Juyeong Ji and Renjie Wu and Jingxiao Zheng and Jyotinder Singh and Abheesht Sharma and Divyashree Sreepathihalli and Aashi Jain and Adham Elarabawy and AJ Co and Andreas Doumanoglou and Babak Samari and Ben Hora and Brian Potetz and Dahun Kim and Enrique Alfonseca and Fedor Moiseev and Feng Han and Frank Palma Gomez and Gustavo Hernández Ábrego and Hesen Zhang and Hui Hui and Jay Han and Karan Gill and Ke Chen and Koert Chen and Madhuri Shanbhogue and Michael Boratko and Paul Suganthan and Sai Meher Karthik Duddu and Sandeep Mariserla and Setareh Ariafar and Shanfeng Zhang and Shijie Zhang and Simon Baumgartner and Sonam Goenka and Steve Qiu and Tanmaya Dabral and Trevor Walker and Vikram Rao and Waleed Khawaja and Wenlei Zhou and Xiaoqi Ren and Ye Xia and Yichang Chen and Yi-Ting Chen and Zhe Dong and Zhongli Ding and Francesco Visin and Gaël Liu and Jiageng Zhang and Kathleen Kenealy and Michelle Casbon and Ravin Kumar and Thomas Mesnard and Zach Gleicher and Cormac Brick and Olivier Lacombe and Adam Roberts and Qin Yin and Yunhsuan Sung and Raphael Hoffmann and Tris Warkentin and Armand Joulin and Tom Duerig and Mojtaba Seyedhosseini},
  eprint = {2509.20354},
  primaryclass = {cs.CL},
  title = {EmbeddingGemma: Powerful and Lightweight Text Representations},
  url = {https://arxiv.org/abs/2509.20354},
  year = {2025}
}

@inproceedings{Jin2025SearchR1,
  author = {Bowen Jin and Hansi Zeng and Zhenrui Yue and Jinsung Yoon and Sercan O Arik and Dong Wang and Hamed Zamani and Jiawei Han},
  booktitle = {COLM 2025},
  title = {Search-R1: Training LLMs to Reason and Leverage Search Engines with Reinforcement Learning},
  url = {https://openreview.net/forum?id=Rwhi91ideu},
  year = {2025}
}

@misc{gu2025surveyllmasajudge,
      title={A Survey on LLM-as-a-Judge}, 
      author={Jiawei Gu and Xuhui Jiang and Zhichao Shi and Hexiang Tan and Xuehao Zhai and Chengjin Xu and Wei Li and Yinghan Shen and Shengjie Ma and Honghao Liu and Saizhuo Wang and Kun Zhang and Yuanzhuo Wang and Wen Gao and Lionel Ni and Jian Guo},
      year={2025},
      eprint={2411.15594},
      archivePrefix={arXiv},
      primaryClass={cs.CL},
      url={https://arxiv.org/abs/2411.15594}, 
}

@misc{yu2024arctic,
  title         = {Arctic-Embed 2.0: Multilingual Retrieval Without Compromise},
  author        = {Puxuan Yu and Luke Merrick and Gaurav Nuti and Daniel Campos},
  year          = {2024},
  eprint        = {2412.04506},
  archiveprefix = {arXiv},
  primaryclass  = {cs.CL},
  url           = {https://arxiv.org/abs/2412.04506}
}

@misc{chen2024bgem3,
  title         = {BGE M3-Embedding: Multi-Lingual, Multi-Functionality, Multi-Granularity Text Embeddings Through Self-Knowledge Distillation},
  author        = {Jianlv Chen and Shitao Xiao and Peitian Zhang and Kun Luo and Defu Lian and Zheng Liu},
  year          = {2024},
  eprint        = {2402.03216},
  archiveprefix = {arXiv},
  primaryclass  = {cs.CL},
  url           = {https://arxiv.org/abs/2402.03216}
}

@misc{wang2024multilinguale5,
  title         = {Multilingual E5 Text Embeddings: A Technical Report},
  author        = {Liang Wang and Nan Yang and Xiaolong Huang and Linjun Yang and Rangan Majumder and Furu Wei},
  year          = {2024},
  eprint        = {2402.05672},
  archiveprefix = {arXiv},
  primaryclass  = {cs.CL},
  url           = {https://arxiv.org/abs/2402.05672}
}

@misc{zhang2024mgte,
  title         = {mGTE: Generalized Long-Context Text Representation and Reranking Models for Multilingual Text Retrieval},
  author        = {Xin Zhang and Yanzhao Zhang and Dingkun Long and Wen Xie and Ziqi Dai and Jialong Tang and Huan Lin and Baosong Yang and Pengjun Xie and Fei Huang and Meishan Zhang and Wenjie Li and Min Zhang},
  year          = {2024},
  eprint        = {2407.19669},
  archiveprefix = {arXiv},
  primaryclass  = {cs.CL},
  url           = {https://arxiv.org/abs/2407.19669}
}

@misc{lee2024mxbai,
  title  = {Open Source Strikes Bread -- New Fluffy Embeddings Model},
  author = {Sean Lee and Aamir Shakir and Darius Koenig and Julius Lipp},
  year   = {2024},
  url    = {https://www.mixedbread.ai/blog/mxbai-embed-large-v1}
}

@misc{li2023angle,
  title         = {AnglE-optimized Text Embeddings},
  author        = {Xianming Li and Jing Li},
  year          = {2023},
  eprint        = {2309.12871},
  archiveprefix = {arXiv},
  primaryclass  = {cs.CL},
  url           = {https://arxiv.org/abs/2309.12871}
}

@misc{nussbaum2024nomic,
  title         = {Nomic Embed: Training a Reproducible Long Context Text Embedder},
  author        = {Zach Nussbaum and John X. Morris and Brandon Duderstadt and Andriy Mulyar},
  year          = {2024},
  eprint        = {2402.01613},
  archiveprefix = {arXiv},
  primaryclass  = {cs.CL},
  url           = {https://arxiv.org/abs/2402.01613}
}

@misc{zhang2025stella,
  title         = {Jasper and Stella: distillation of SOTA embedding models},
  author        = {Dun Zhang and Jiacheng Li and Ziyang Zeng and Fulong Wang},
  year          = {2025},
  eprint        = {2412.19048},
  archiveprefix = {arXiv},
  primaryclass  = {cs.IR},
  url           = {https://arxiv.org/abs/2412.19048}
}

@article{robertson2009bm25,
  title   = {The Probabilistic Relevance Framework: BM25 and Beyond},
  author  = {Stephen Robertson and Hugo Zaragoza},
  journal = {Foundations and Trends in Information Retrieval},
  volume  = {3},
  number  = {4},
  pages   = {333--389},
  year    = {2009},
  doi     = {10.1561/1500000019}
}

@inproceedings{santhanam2022colbertv2,
  title     = {ColBERTv2: Effective and Efficient Retrieval via Lightweight Late Interaction},
  author    = {Keshav Santhanam and Omar Khattab and Jon Saad-Falcon and Christopher Potts and Matei Zaharia},
  booktitle = {Proceedings of the 2022 Conference of the North American Chapter of the Association for Computational Linguistics: Human Language Technologies},
  year      = {2022},
  url       = {https://aclanthology.org/2022.naacl-main.272}
}

@inproceedings{Cormack2009RRF,
author = {Cormack, Gordon V. and Clarke, Charles L A and Buettcher, Stefan},
title = {Reciprocal rank fusion outperforms condorcet and individual rank learning methods},
year = {2009},
isbn = {9781605584836},
publisher = {Association for Computing Machinery},
address = {New York, NY, USA},
url = {https://doi.org/10.1145/1571941.1572114},
doi = {10.1145/1571941.1572114},
booktitle = {Proceedings of the 32nd International ACM SIGIR Conference on Research and Development in Information Retrieval},
pages = {758–759},
numpages = {2},
location = {Boston, MA, USA},
series = {SIGIR '09}
}

@inproceedings{eslami-etal-2026-diffusion,
    title = "Diffusion-Pretrained Dense and Contextual Embeddings",
    author = "Eslami, Sedigheh  and
      Gaiduk, Maksim  and
      Krimmel, Markus  and
      Milliken, Louis Mark  and
      Wang, Bo  and
      Bykov, Denis",
    editor = "Li, Yunyao  and
      Rehm, Georg  and
      Tu, Mei",
    booktitle = "Proceedings of the 64th Annual Meeting of the {A}ssociation for {C}omputational {L}inguistics (Volume 6: Industry Track)",
    month = jul,
    year = "2026",
    address = "San Diego, California, USA",
    publisher = "Association for Computational Linguistics",
    url = "https://aclanthology.org/2026.acl-industry.69/",
    doi = "10.18653/v1/2026.acl-industry.69",
    pages = "990--1004",
    ISBN = "979-8-89176-394-4"
}

@article{Tsatsaronis2015BioASQ,
  title     = {An Overview of the {BIOASQ} Large-Scale Biomedical Semantic Indexing and Question Answering Competition},
  author    = {Tsatsaronis, George and Balikas, Georgios and Malakasiotis, Prodromos and Partalas, Ioannis and Zschunke, Matthias and Alvers, Michael R. and Weissenborn, Dirk and Krithara, Anastasia and Petridis, Sergios and Polychronopoulos, Dimitris and Almirantis, Yannis and Pavlopoulos, John and Baskiotis, Nicolas and Gallinari, Patrick and Arti{\`e}res, Thierry and Ngonga Ngomo, Axel-Cyrille and Heino, Norman and Gaussier, {\'E}ric and Barrio-Alvers, Liliana and Schroeder, Michael and Androutsopoulos, Ion and Paliouras, Georgios},
  journal   = {BMC Bioinformatics},
  volume    = {16},
  number    = {1},
  pages     = {138},
  year      = {2015},
  doi       = {10.1186/s12859-015-0564-6}
}

@inproceedings{Thorne2018FEVER,
  title     = {{FEVER}: A Large-Scale Dataset for Fact Extraction and {VERification}},
  author    = {Thorne, James and Vlachos, Andreas and Christodoulopoulos, Christos and Mittal, Arpit},
  booktitle = {Proceedings of the 2018 Conference of the North American Chapter of the Association for Computational Linguistics: Human Language Technologies (NAACL-HLT)},
  pages     = {809--819},
  year      = {2018},
  doi       = {10.18653/v1/N18-1074}
}

@inproceedings{Yang2018HotpotQA,
  title     = {{HotpotQA}: A Dataset for Diverse, Explainable Multi-Hop Question Answering},
  author    = {Yang, Zhilin and Qi, Peng and Zhang, Saizheng and Bengio, Yoshua and Cohen, William W. and Salakhutdinov, Ruslan and Manning, Christopher D.},
  booktitle = {Proceedings of the 2018 Conference on Empirical Methods in Natural Language Processing (EMNLP)},
  pages     = {2369--2380},
  year      = {2018},
  doi       = {10.18653/v1/D18-1259}
}

@inproceedings{Voorhees2002Reliability,
  title     = {The Effect of Topic Set Size on Retrieval Experiment Error},
  author    = {Voorhees, Ellen M. and Buckley, Chris},
  booktitle = {Proceedings of the 25th Annual International {ACM} {SIGIR} Conference on Research and Development in Information Retrieval},
  pages     = {316--323},
  year      = {2002},
  doi       = {10.1145/564376.564432}
}

@inproceedings{Qu2021RocketQA,
  title     = {{RocketQA}: An Optimized Training Approach to Dense Passage Retrieval for Open-Domain Question Answering},
  author    = {Qu, Yingqi and Ding, Yuchen and Liu, Jing and Liu, Kai and Ren, Ruiyang and Zhao, Wayne Xin and Dong, Daxiang and Wu, Hua and Wang, Haifeng},
  booktitle = {Proceedings of the 2021 Conference of the North American Chapter of the Association for Computational Linguistics: Human Language Technologies (NAACL-HLT)},
  pages     = {5835--5847},
  year      = {2021},
  doi       = {10.18653/v1/2021.naacl-main.466}
}

@article{Arabzadeh2022ShallowPooling,
  title     = {Shallow Pooling for Sparse Labels},
  author    = {Arabzadeh, Negar and Vtyurina, Alexandra and Yan, Xinyi and Clarke, Charles L. A.},
  journal   = {Information Retrieval Journal},
  volume    = {25},
  pages     = {365--385},
  year      = {2022},
  doi       = {10.1007/s10791-022-09411-0}
}

@inproceedings{Buckley2004Incomplete,
  title     = {Retrieval Evaluation with Incomplete Information},
  author    = {Buckley, Chris and Voorhees, Ellen M.},
  booktitle = {Proceedings of the 27th Annual International {ACM} {SIGIR} Conference on Research and Development in Information Retrieval},
  pages     = {25--32},
  year      = {2004},
  doi       = {10.1145/1008992.1009000}
}

@inproceedings{Wang2011Cascade,
  author = {Lidan Wang and Jimmy Lin and Donald Metzler},
  title = {A Cascade Ranking Model for Efficient Ranked Retrieval},
  booktitle = {Proceedings of SIGIR},
  pages = {105--114},
  year = {2011}
}

@article{Nogueira2019MultiStage,
  author = {Rodrigo Nogueira and Wei Yang and Kyunghyun Cho and Jimmy Lin},
  title = {Multi-Stage Document Ranking with {BERT}},
  journal = {arXiv:1910.14424},
  year = {2019}
}

@inproceedings{Sainz2023Contamination,
  title     = {{NLP} Evaluation in trouble: On the Need to Measure {LLM} Data Contamination for each Benchmark},
  author    = {Sainz, Oscar and Campos, Jon Ander and Garc{\'i}a-Ferrero, Iker and Etxaniz, Julen and Lopez de Lacalle, Oier and Agirre, Eneko},
  booktitle = {Findings of the Association for Computational Linguistics: EMNLP 2023},
  year      = {2023},
  address   = {Singapore},
  publisher = {Association for Computational Linguistics},
  pages     = {10776--10787},
  doi       = {10.18653/v1/2023.findings-emnlp.722},
  url       = {https://aclanthology.org/2023.findings-emnlp.722/}
}

@inproceedings{Matton2024CodeLeakage,
  title     = {On Leakage of Code Generation Evaluation Datasets},
  author    = {Matton, Alexandre and Sherborne, Tom and Aumiller, Dennis and Tommasone, Elena and Alizadeh, Milad and He, Jingyi and Ma, Raymond and Voisin, Maxime and Gilsenan-McMahon, Ellen and Gall{\'e}, Matthias},
  booktitle = {Findings of the Association for Computational Linguistics: EMNLP 2024},
  year      = {2024},
  address   = {Miami, Florida, USA},
  publisher = {Association for Computational Linguistics},
  pages     = {13215--13223},
  doi       = {10.18653/v1/2024.findings-emnlp.772},
  url       = {https://aclanthology.org/2024.findings-emnlp.772/}
}

@inproceedings{Oren2024Proving,
  title     = {Proving Test Set Contamination in Black Box Language Models},
  author    = {Oren, Yonatan and Meister, Nicole and Chatterji, Niladri S. and Ladhak, Faisal and Hashimoto, Tatsunori B.},
  booktitle = {The Twelfth International Conference on Learning Representations (ICLR)},
  year      = {2024},
  url       = {https://openreview.net/forum?id=KS8mIvetg2}
}

@inproceedings{Reimers2021Curse,
  title     = {The Curse of Dense Low-Dimensional Information Retrieval for Large Index Sizes},
  author    = {Reimers, Nils and Gurevych, Iryna},
  booktitle = {Proceedings of the 59th Annual Meeting of the Association for Computational Linguistics and the 11th International Joint Conference on Natural Language Processing (Volume 2: Short Papers)},
  year      = {2021},
  pages     = {605--611},
  publisher = {Association for Computational Linguistics},
  url       = {https://aclanthology.org/2021.acl-short.77/}
}

@article{buckley2007bias,
  author  = {Chris Buckley and Darrin Dimmick and Ian Soboroff and Ellen Voorhees},
  title   = {Bias and the Limits of Pooling for Large Collections},
  journal = {Information Retrieval},
  volume  = {10},
  number  = {6},
  pages   = {491--508},
  year    = {2007},
  doi     = {10.1007/s10791-007-9032-x}
}

@inproceedings{DBLP:conf/ecir/CaspariDDFMG26,
  author       = {Laura Caspari and
                  Michael Dinzinger and
                  Kanishka Ghosh Dastidar and
                  Christofer Fellicious and
                  Jelena Mitrovic and
                  Michael Granitzer},
  editor       = {Ricardo Campos and
                  Adam Jatowt and
                  Yanyan Lan and
                  Mohammad Aliannejadi and
                  Christine Bauer and
                  Sean MacAvaney and
                  Avishek Anand and
                  Zhaochun Ren and
                  Suzan Verberne and
                  Nan Bai and
                  Masoud Mansoury},
  title        = {CoRECT: {A} Framework for Evaluating Embedding Compression Techniques
                  at Scale},
  booktitle    = {Advances in Information Retrieval - 48th European Conference on Information
                  Retrieval, {ECIR} 2026, Delft, The Netherlands, March 29 - April 2,
                  2026, Proceedings, Part {IV}},
  series       = {Lecture Notes in Computer Science},
  volume       = {16486},
  pages        = {383--398},
  publisher    = {Springer},
  year         = {2026},
  url          = {https://doi.org/10.1007/978-3-032-21321-1_48},
  doi          = {10.1007/978-3-032-21321-1_48},
  bibsource    = {dblp computer science bibliography, https://dblp.org}
}

@inproceedings{DBLP:conf/ecir/FrobeSRHMHLNVP26,
  author       = {Maik Fr{\"{o}}be and
                  Ferdinand Schlatt and
                  Cosimo Rulli and
                  Tim Hagen and
                  Jan Heinrich Merker and
                  Gijs Hendriksen and
                  Carlos Lassance and
                  Franco Maria Nardini and
                  Rossano Venturini and
                  Martin Potthast},
  editor       = {Ricardo Campos and
                  Adam Jatowt and
                  Yanyan Lan and
                  Mohammad Aliannejadi and
                  Christine Bauer and
                  Sean MacAvaney and
                  Avishek Anand and
                  Zhaochun Ren and
                  Suzan Verberne and
                  Nan Bai and
                  Masoud Mansoury},
  title        = {Evaluating the Efficiency and Effectiveness of Learned Sparse Retrieval
                  with the lsr{\_}benchmark},
  booktitle    = {Advances in Information Retrieval - 48th European Conference on Information
                  Retrieval, {ECIR} 2026, Delft, The Netherlands, March 29 - April 2,
                  2026, Proceedings, Part {IV}},
  series       = {Lecture Notes in Computer Science},
  volume       = {16486},
  pages        = {528--543},
  publisher    = {Springer},
  year         = {2026},
  url          = {https://doi.org/10.1007/978-3-032-21321-1_57},
  doi          = {10.1007/978-3-032-21321-1_57},
  bibsource    = {dblp computer science bibliography, https://dblp.org}
}

@article{Wei2024SimpleQA,
  author       = {Jason Wei and
                  Nguyen Karina and
                  Hyung Won Chung and
                  Yunxin Joy Jiao and
                  Spencer Papay and
                  Amelia Glaese and
                  John Schulman and
                  William Fedus},
  title        = {Measuring short-form factuality in large language models},
  journal      = {CoRR},
  volume       = {abs/2411.04368},
  year         = {2024},
  url          = {https://doi.org/10.48550/arXiv.2411.04368},
  doi          = {10.48550/ARXIV.2411.04368},
  eprinttype   = {arXiv},
  eprint       = {2411.04368},
  bibsource    = {dblp computer science bibliography, https://dblp.org}
}

@article{DBLP:journals/corr/abs-2504-12516,
  author       = {Jason Wei and
                  Zhiqing Sun and
                  Spencer Papay and
                  Scott McKinney and
                  Jeffrey Han and
                  Isa Fulford and
                  Hyung Won Chung and
                  Alex Tachard Passos and
                  William Fedus and
                  Amelia Glaese},
  title        = {BrowseComp: {A} Simple Yet Challenging Benchmark for Browsing Agents},
  journal      = {CoRR},
  volume       = {abs/2504.12516},
  year         = {2025},
  url          = {https://doi.org/10.48550/arXiv.2504.12516},
  doi          = {10.48550/ARXIV.2504.12516},
  eprinttype   = {arXiv},
  eprint       = {2504.12516},
  bibsource    = {dblp computer science bibliography, https://dblp.org}
}

@article{Formal2021SPLADEVS,
  title={SPLADE v2: Sparse Lexical and Expansion Model for Information Retrieval},
  author={Thibault Formal and Carlos Lassance and Benjamin Piwowarski and St{\'e}phane Clinchant},
  journal={ArXiv},
  year={2021},
  volume={abs/2109.10086},
  url={https://api.semanticscholar.org/CorpusID:237581550}
}

@article{kendall1945treatment,
  author  = {Kendall, Maurice G.},
  title   = {The Treatment of Ties in Ranking Problems},
  journal = {Biometrika},
  volume  = {33},
  number  = {3},
  pages   = {239--251},
  year    = {1945},
  doi     = {10.1093/biomet/33.3.239}
}

@inproceedings{craswell2021trec,
    author = {Nick Craswell and Bhaskar Mitra and Emine Yilmaz and Daniel Campos and Jimmy Lin},
    editor = {Ian Soboroff and Angela Ellis},
    title = {Overview of the {TREC} 2021 Deep Learning Track},
    booktitle = {Proceedings of the Thirtieth Text REtrieval Conference, {TREC} 2021, online, November 15-19, 2021},
    series = {{NIST} Special Publication},
    volume = {500-335},
    publisher = {National Institute of Standards and Technology {(NIST)}},
    year = {2021},
    url = {https://trec.nist.gov/pubs/trec30/papers/Overview-DL.pdf},
    bibsource = {dblp computer science bibliography, https://dblp.org},
}

@inproceedings{Penha2022QueryVariations,
author = {Penha, Gustavo and C{\^a}mara, Arthur and Hauff, Claudia},
title = {Evaluating the Robustness of Retrieval Pipelines with Query Variation Generators},
year = {2022},
isbn = {978-3-030-99735-9},
publisher = {Springer-Verlag},
address = {Berlin, Heidelberg},
url = {https://doi.org/10.1007/978-3-030-99736-6_27},
doi = {10.1007/978-3-030-99736-6_27},
booktitle = {Advances in Information Retrieval: 44th European Conference on IR Research, ECIR 2022, Stavanger, Norway, April 10–14, 2022, Proceedings, Part I},
pages = {397–412},
numpages = {16},
location = {Stavanger, Norway}
}
\clearpage
\appendix
\section{Retrievers for Relevance Pooling and Subcorpus Sampling}
\label{app:pooling-models}

\begin{table}[h]
\caption{Retrievers used to pool previously unjudged documents for the
\textsc{Combined+LLM-Judged} pass. All were released before the January 1, 2025 cutoff,
and none of the dense or late-interaction models is among the thirteen retrievers
evaluated in \autoref{sec:benchmark-retrievers}. Dim.\ is the embedding dimension,
per token for late interaction; stella-400m-v5 and arctic-l-v2 also support smaller
dimensions. Max tok.\ is the input length we index with.}
\label{tab:pooling-models}
\centering
\footnotesize
\setlength{\tabcolsep}{5pt}
\begin{tabular}{@{}l l r r r l@{}}
\toprule
\textbf{Model} & \textbf{Type} & \textbf{Params} & \textbf{Dim.} &
\textbf{Max tok.} & \textbf{Lang.} \\
\midrule
BM25-tantivy~\citep{robertson2009bm25}                   & lexical   & --   & --   & --   & multi \\
ColBERTv2~\citep{santhanam2022colbertv2}         & late int. & 110M & 128  & 180  & en \\
arctic-l-v2~\citep{yu2024arctic}                 & dense     & 568M & 1024 & 8192 & multi \\
bge-m3~\citep{chen2024bgem3}                     & dense     & 568M & 1024 & 8192 & multi \\
e5-large-instruct~\citep{wang2024multilinguale5} & dense     & 560M & 1024 & 512  & multi \\
gte-multilingual-base~\citep{zhang2024mgte}      & dense     & 305M & 768  & 8192 & multi \\
mxbai-large-v1~\citep{lee2024mxbai,li2023angle}  & dense     & 335M & 1024 & 512  & en \\
nomic-v1.5~\citep{nussbaum2024nomic}             & dense     & 137M & 768  & 8192 & en \\
stella-400m-v5~\citep{zhang2025stella}           & dense     & 435M & 1024 & 512  & en \\
\bottomrule
\end{tabular}
\end{table}

\end{document}